\documentclass[aps,prd,a4paper,twocolumn,nofootinbib,superscriptaddress,10pt]{revtex4-1}

\usepackage{amsmath, amsfonts, amsthm, amssymb, graphicx, color, hyperref}
\usepackage{subfigure}

\usepackage[utf8]{inputenc}

\allowdisplaybreaks[3]

\def \bal#1\eal  {\begin{align} #1 \end{align}}
\def\({\left(}
\def\){\right)}
\def\[{\left[}
\def\]{\right]}
\def\<{\left\langle}
\def\>{\right\rangle}
\def\d{\mathrm{d}}

\newcommand{\f}[2]{\frac{#1}{#2}}
\newcommand{\bim} {\begin{itemize}[noitemsep]}

\newcommand{\eim}{\end{itemize}}
\newcommand{\be} {\begin{equation}}
\newcommand{\ee} {\end{equation}}
\newcommand{\bc}{\begin{center}}
\newcommand{\ec}{\end{center}}

\newcommand{\nn} {\nonumber\\}

\newcommand{\mc} {\mathcal}

\newcommand{\ai}{{\alpha}}
\newcommand{\bi}{{\beta}}

\newcommand{\ri}{{\rho}}
\newcommand{\si}{{\sigma}}
\newcommand{\li}{{\lambda}}
\newcommand{\ti}{{\tau}}

\newcommand{\thi}{\theta}

\begin{document}

\hfill {\footnotesize USTC-ICTS-19-29}

\title{Relativistic stars in mass-varying massive gravity}

\author{Xue Sun}
\affiliation{School of Physical Sciences, University of Science and Technology of China, Hefei, Anhui 230026, China}

\author{Shuang-Yong Zhou}
\affiliation{Interdisciplinary Center for Theoretical Study, University of Science and Technology of China, Hefei, Anhui 230026, China\\
and Peng Huanwu Center for Fundamental Theory, Hefei, Anhui 230026, China}

\date{\today}

\begin{abstract}

Mass-varying massive gravity allows the graviton mass to vary according to different environments. We investigate neutron star and white dwarf solutions in this theory and find that the graviton mass can become very large near the compact stars and settle down quickly to small cosmological values away the stars, similar to that of black holes in the theory. It is found that there exists a tower of compact star solutions where the graviton mass decreases radially to zero non-trivially. We compute the massive graviton effects on the mass-radius relations of the compact stars, and also compare the relative strengths between neutron stars and white dwarfs in constraining the parameter space of mass-varying massive gravity.\\
~\\

\end{abstract}

\maketitle
\flushbottom

\section{Introduction}

Recent advances in astrophysics, particularly the arrival of gravitational wave astronomy \cite{Abbott:2016blz, TheLIGOScientific:2017qsa}, have provided fresh new opportunities to test gravity in the strong field regime with compact astronomical objects. Whilst in Einstein gravity the graviton is massless, an interesting class of alternative theories of gravity is to let the graviton become massive \cite{deRham:2010ik, deRham:2010kj, Hassan:2011hr, Hassan:2011ea}. See \cite{deRham:2014zqa, Schmidt-May:2015vnx, Hinterbichler:2011tt} for a recent review on recent developments in constructing massive gravity and bi-gravity models \cite{deRham:2010ik, deRham:2010kj, Hassan:2011hr, Hassan:2011ea} that are free of the Boulware-Deser ghost \cite{Boulware:1973my} and their applications. Typically, the graviton mass is set to be close to the current Hubble scale \cite{deRham:2016nuf}, as one of the main phenomenological applications of these models is to explain the late time cosmic acceleration \cite{Riess:1998cb, Perlmutter:1998np}. For such a small graviton mass, or such a long Compton wavelength, simple dimension analysis suggests that these models can be best constrained by large scale or cosmological observations, although sometimes the solar system gravity tests may achieve comparable bounds, thanks to their superb accuracy \cite{deRham:2016nuf}.

However, the graviton mass does not have to be uniformly small across the whole spacetime. In mass-varying massive gravity \cite{Huang:2012pe, Huang:2013mha}, the original de Rham-Gabadadze-Tolley (dRGT) model \cite{deRham:2010kj} is augmented with an extra environmental scalar field and the graviton mass now depends on the environmental field, which can take different values in various different astrophysical environments. In \cite{Zhang:2017jze}, it is shown that the graviton mass can become extremely large near the event horizon of the hairy black hole in mass-varying massive gravity \cite{Tolley:2015ywa}, while setting down to the small cosmological value away from the horizon, thus satisfying all the current tests of gravity and yet still giving rise to interesting deviations from general relativity that are testable in the current and upcoming experiments. An important feature of those hairy black holes is that because of the rapid increase of the graviton mass near the horizon, there is an extra potential barrier to the left of the photosphere barrier in the modified Regge-Wheeler-Zerilli equation, and this can lead to gravitational wave echoes \cite{Cardoso:2016rao, Cardoso:2017cqb} in the late time ringdown waveform when the black hole is perturbed \cite{Zhang:2017jze}.

The detection of gravitational waves from neutron star mergers \cite{TheLIGOScientific:2017qsa} has added a new dimension to our ability to unravel the nature of strong gravity and the structure of relativistic compact stars. Neutron star solutions have been previously investigated in the original dRGT model \cite{Katsuragawa:2015lbl} and in bi-gravity \cite{Sullivan:2017kwo} with a graviton mass that is much greater than typical cosmological scales. Relativistic star solutions in dRGT-like models with a singular reference metric have also been studied \cite{Hendi:2017ibm, EslamPanah:2018evk}.

In this paper, we will further investigate the compact star solutions in mass-varying massive gravity. We will compute the neutron star and white dwarf solutions, using the APR \cite{Akmal:1998cf} and SLy \cite{Douchin:2001sv} equations of state for neutron stars and the Chandrasekhar equation of state for white dwarfs, and analyze the behavior of the graviton mass near compact stars. We will study how the effects of a varying graviton mass affect the mass-radius diagram of compact stars and compare the relative strengths of neutron stars and white dwarfs in constraining the parameter space of mass-varying massive gravity.

The paper is organized as follows: We introduce mass-varying massive gravity in Section \ref{sec:mvmg}, and in Section \ref{sec:setup}, we reduce the field equations into a system of ordinary differential equations for compact stars and describe our numerical setup; In Section \ref{sec:ns}, we set up the equations of state for neutron stars, compute their solutions and investigate how the mass-radius diagrams change with the model parameters; In Section \ref{sec:wd}, we investigate the white dwarf solutions and compare the relative strengths of neutron stars and white dwarfs in constraining the parameters in mass-varying massive gravity; We conclude in Section \ref{conclu}. We shall use geometric units which set $c=G=1$ and express dimensionful quantities in terms of kilometers.

\section{Mass-varying massive gravity}
\label{sec:mvmg}

Mass-varying massive gravity (MVMG) \cite{Huang:2012pe} is extended dRGT massive gravity \cite{deRham:2010kj} where the mass of the graviton can vary according to the value of a scalar field in different environments. Its action is given by
\bal
S &= \f1{ 8\pi } \int \d^4 x \sqrt{-g}\bigg[\frac{\cal R}{2}+V(\sigma) U(\mathcal{K})
\nn
&~~~~~~~~~-\frac{1}{2} g^{\mu \nu} \partial_{\mu} \sigma \partial_{\nu} \sigma-W(\sigma)\bigg]+S_{\text{m}}\left[g_{\mu \nu}\right] ,
\eal
where $S_{\text{m}}$ is the action of the conventional matter which only couples to the dynamic metric $g_{\mu\nu}$ and the dRGT graviton potential is given by
\be
U(\mathcal{K}) =  \mathcal{K}_{\left[\mu\right.}^{\mu} \mathcal{K}_{\left.\nu\right]}^{\nu}+\ai_3 \mathcal{K}_{\left[\mu\right.}^{\mu} \mathcal{K}_{\nu}^{\nu} \mathcal{K}_{\left.\ri\right]}^{\ri} + \ai_4 \mathcal{K}_{\left[\mu\right.}^{\mu} \mathcal{K}_{\nu}^{\nu}\mathcal{K}_{\ri}^{\ri} \mathcal{K}_{\left.\si\right]}^{\si}  ,
\ee
with $\mathcal{K}_{\nu}^{\mu} \equiv \delta_{\nu}^{\mu}-\sqrt{g^{-1} \eta}\big|_{\nu} ^{\mu}$, $g^{-1}=\left(g^{\mu \nu}\right)$ being the inverse metric and $\eta=\left(\eta_{\mu \nu}\right)$ being the Minkowski metric. The anti-symmetrization is defined with unit weight, {\it e.g.}, $\mathcal{K}_{\left[\mu\right.}^{\mu} \mathcal{K}_{\left.\nu\right]}^{\nu}= (\mathcal{K}_{\mu}^{\mu} \mathcal{K}_{\nu}^{\nu}-\mathcal{K}_{\nu}^{\mu} \mathcal{K}_{\mu}^{\nu})/2$. There are of course many choices for $V(\si)$ and $W(\si)$, which are functions of the environment scalar field $\si$. We consider a simple model where these potentials are \cite{Zhang:2017jze}
\be
\label{VWform}
V(\sigma)=m_{0}^{2}+m^{2} \sigma^{4}, \quad W(\sigma)=\frac{1}{2} m_{\sigma}^{2} \sigma^{2}+\lambda_{\sigma} \sigma^{6},
\ee
where $m_0$ is chosen to be at most of the order of the Hubble scale, potentially generating the late time cosmic acceleration. For star solutions, which are well within the Hubble horizon, we can safely neglect the $m_{0}^{2}$ term. The $\sigma^{6}$ term, where $\li_\si$ is small, is added for the stability of the vacuum and can be neglected for local star solutions \cite{Zhang:2017jze}. So apart from $\ai_3$ and $\ai_4$, the essential theory parameters are $m$ and $m_\si$, the dependence of which will be explored for the compact stars in the next sections.

The equations of motion for this model are given by
\bal
G_{\mu \nu} &=8 \pi   T_{\mu \nu}+T^{(\si)}_{\mu \nu}+V(\sigma) X_{\mu \nu}  ,
\\
\partial_{\mu}\left(\sqrt{-g} g^{\mu v} \partial_{v} \sigma\right)&=\sqrt{-g}\left(W_{\sigma}-V_{\sigma} U\right) ,
\eal
where $V_\si= \d V /\d \si$, $W_\si = \d W/\d \si$, $T_{\mu\nu}$ is the energy momentum tensor from $S_{\rm m}$,
\be
T^{(\si)}_{\mu \nu}= \partial_{\mu} \sigma \partial_{\nu} \sigma-g_{\mu \nu}\left(\frac{1}{2} g^{\rho \gamma} \partial_{\rho} \sigma \partial_{\gamma} \sigma+W(\sigma)\right)
\ee
 and
\bal
X_{\mu \nu}&= -\left( g_{\mu\ri} \mathcal{K}^{\ri}_{\nu}-\mathcal{K}^{\mu_1}_{\mu_1} g_{\mu \nu}\right)
\nn
&~~~+\alpha\left(  g_{\mu\ri}\mathcal{K}^{\ri}_{\si} \mathcal{K}^{\si}_{\nu} -\mathcal{K}^{\mu_1}_{\mu_1} g_{\mu\ri} \mathcal{K}^{\ri}_{\nu}+\mathcal{K}_{\left[\mu_1\right.}^{\mu_1} \mathcal{K}_{\left.\mu_2\right]}^{\mu_2} g_{\mu \nu}\right)
\nn
&~~~ -\beta\Big(g_{\mu\ri} \mathcal{K}^{\ri}_{\si} \mathcal{K}^{\si}_{\ti} \mathcal{K}^{\ti}_{\nu} -\mathcal{K}^{\mu_1}_{\mu_1} g_{\mu\ri}\mathcal{K}^{\ri}_{\si} \mathcal{K}^{\si}_{\nu}
\nn
&~~~ +\mathcal{K}_{\left[\mu_1\right.}^{\mu_1} \mathcal{K}_{\left.\mu_2\right]}^{\mu_2}  g_{\mu\ri} \mathcal{K}^{\ri}_{\nu} - \mathcal{K}_{\left[\mu_1\right.}^{\mu_1} \mathcal{K}_{\mu_2}^{\mu_2} \mathcal{K}_{\left.\mu_3\right]}^{\mu_3} g_{\mu \nu}\Big)  .
\eal
$\ai$ and $\bi$ are related to $\ai_3$ and $\ai_4$ by
\be
 \alpha=1+\alpha_{3}, \quad \beta=\alpha_{3}+\alpha_{4}  .
\ee

\section{Setup}
\label{sec:setup}

We consider a static and spherically symmetric Ansatz for the scalar field $\si=\si(r)$ and
for the metric
\bal
\label{dsorigin}
\mathrm{d} s^{2}&=-a(r) \mathrm{d}t^{2}+2 b(r) \mathrm{d}r \mathrm{d}t+e(r) \mathrm{d}r^{2}+d(r) \mathrm{d} \Omega^{2}  ,
\\
\mathrm{d} \mathrm{s}_{\eta}^{2}&=-\mathrm{d}t^{2}+\mathrm{d}r^{2}+r^{2} \mathrm{d} \Omega^{2}  .
\eal
Plugging these into the Einstein tensor, the nonzero components are
\bal
G_{t}^{t}&=\frac{2 a^{\prime} c d d^{\prime}+4 a c d d^{\prime \prime} -4 c^{2} d-2 a c^{\prime} d d^{\prime}-a c d^{\prime 2}}{4 c^{2} d^{2}}  ,
\\
G_{r}^{t}&=\frac{b c^{\prime} d d^{\prime}+b c d^{\prime 2}-2 b c d d^{\prime \prime}}{2 c^{2} d^{2}}  ,
\\
G_{r}^{r}&=\frac{-4 c d+2 a^{\prime} d d^{\prime}+a d^{\prime 2}}{4 c d^{2}}  ,
\\
G_{\theta}^{\theta}&=\frac{1}{4 c^{2} d^{2}}  \big(-a^{\prime} c^{\prime} d^{2}+2 c a^{\prime} d d^{\prime}-a c^{\prime} d d^{\prime}-a c d^{\prime 2}
\nn
&~~~ +2 a^{\prime \prime} c d^{2}+2 a c d d^{\prime \prime} \big)  ,
\eal
where $'$ denotes a derivative with respective to the radius $r$ and by symmetry we have $G_{\phi}^{\phi}=G_{\theta}^{\theta}$. The matter energy momentum tensor is taken to be a perfect fluid
\be
{T}_{\mu\nu}=(\rho(r) + p(r)) u_{\mu} u_{\nu}+p(r) g_{\mu\nu}   ,
\ee
where $\ri$ is the matter energy density, $p$ is the matter pressure and $u^\mu$ is the 4-velocity of the perfect fluid, which, in the static case, is determined completely by the normalization $g_{\mu\nu}u^\mu u^\nu =-1$. Then the non-trivial components of the matter energy momentum tensor are
\be
{T}_{t}^{t}=- \rho,~~~{T}_{r}^{t}=\frac{(p+\rho) b}{a},~~~{T}_{r}^{r}={T}_{\theta}^{\theta}={T}_{\phi}^{\phi}= p   .
\ee
On the other hand, the nonzero components of the energy-momentum tensor generated by the scalar $\si$ are
\be
T^{(\si)}{}_{t}^{t}=T^{(\si)}{}_{\theta}^{\theta}=T^{(\si)}{}_{\phi}^{\phi}=-W(\sigma)-\frac{a \sigma^{\prime 2}}{2 c}  ,
\ee
and the nonzero components of the effective energy-momentum tensor from the massive graviton are
\bal
X_{t}^{t}&=\frac{\left(b^{2} n^{2} k_{1}-k_{2}\right)K_3}{b^{2} n^{2}-1}+k_{3}\left(\alpha k_{3}+2\right)  ,
\\
X_{r}^{t} &=-X_{t}^{r}=\frac{bK_3}{c}  ,
\\
X_{r}^{r} &=\frac{\left(b^{2} n^{2} k_{2}-k_{1}\right)K_3}{b^{2} n^{2}-1}+k_{3}\left(\alpha k_{3}+2\right)   ,
\\
X_{\theta}^{\theta}&=X_{\phi}^{\phi}=\alpha\left(k_{1} k_{2}+k_{1} k_{3}+k_{2} k_{3}\right)+\beta k_{1} k_{2} k_{3}
\nn
& ~~~~~~~~~~~ +k_{1}+k_{2}+k_{3}   ,
\eal
where $K_3= \beta k_{3}^{2}+2 \alpha k_{3}+1$ and $k_1$, $k_2$ and $k_3$ are the eigenvalues of matrix $\mc{K}^\mu_\nu$, explicitly given by
\bal
k_{1}&=1-\sqrt{\frac{2}{a+e+\sqrt{(a+e)^{2}-4 c}}}  ,
\\
k_{2}&=1-\sqrt{\frac{2}{a+e-\sqrt{(a+e)^{2}-4 c}}}  ,
\\
\label{k3equ}
k_{3}&=1-\sqrt{\frac{r^{2}}{d}}  ,
\eal
with $n= {2}/{(\sqrt{(a+e)^{2}-4 c}+a-e)}$ and $c= b^{2}+a e$.

\begin{figure}
\centering
\includegraphics[width=0.4\textwidth]{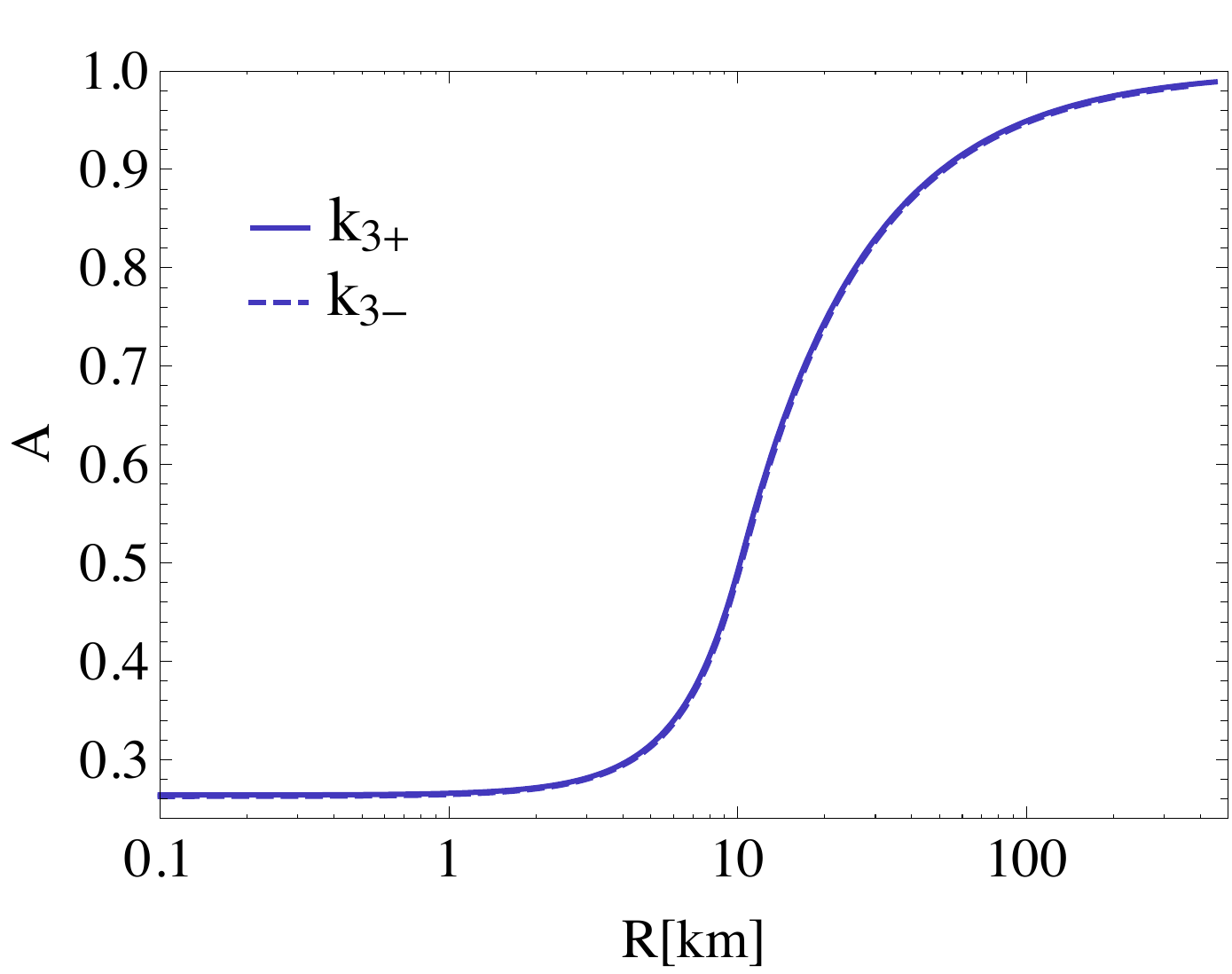}
\includegraphics[width=0.4\textwidth]{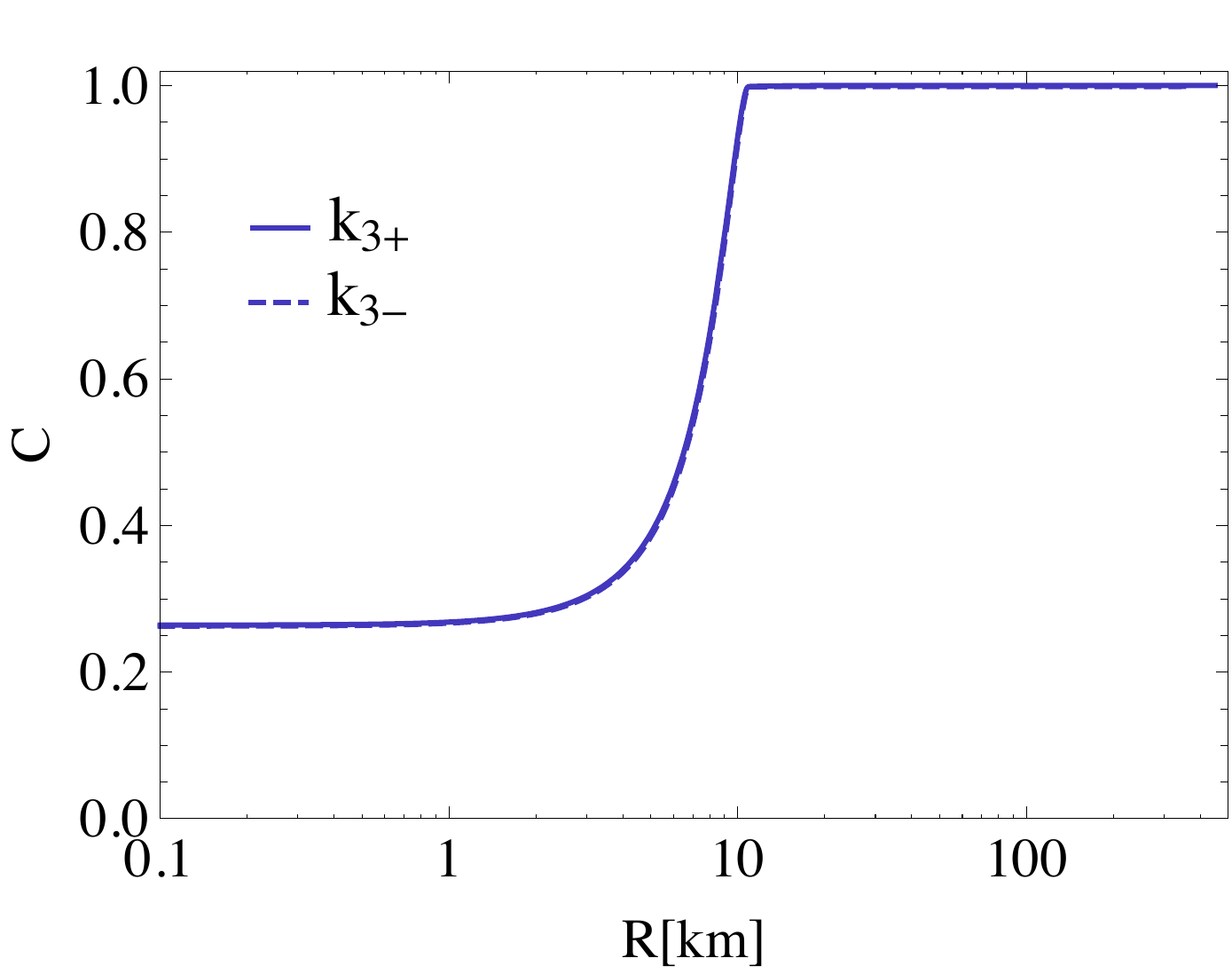}
\includegraphics[width=0.4\textwidth]{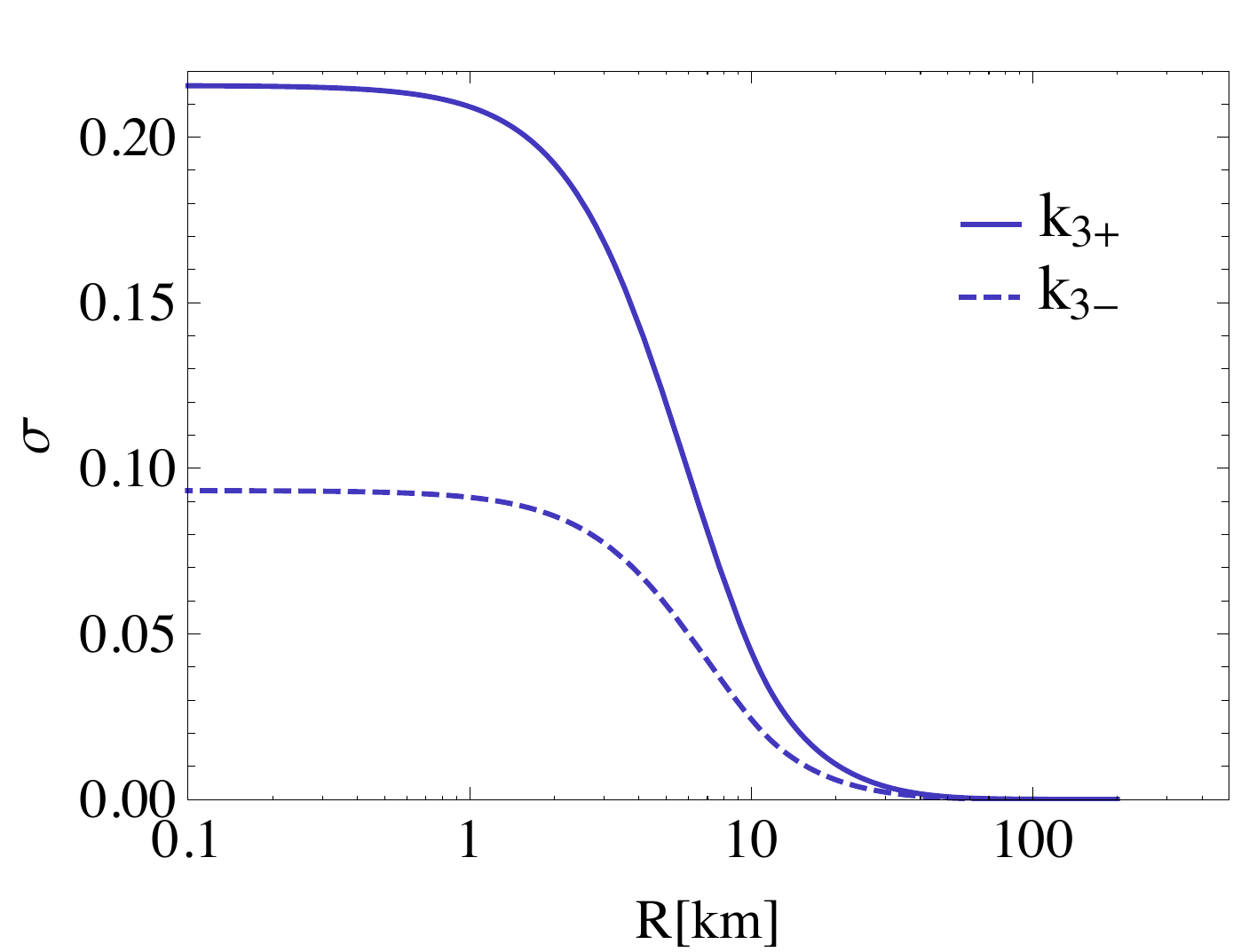}
\caption{
Radial profiles of representative neutron stars. $A$ is the ${}_{tt}$ metric component, $C=B^2+AE$ (see metric (\ref{dsrescale})) and $\sigma$ is the environmental scalar field. The graviton potential parameters are $\alpha=2, \beta=3$, which leads to ${k}_{3+}=-1 / 3$ and  ${k}_{3-}=-1$ for the two branches of solutions. The two branches of solutions are only slightly different from each other in the top two plots. We use geometric units $G=c=1$ and express dimensionful quantities in kilometers. The mass parameters are chosen as $\mathrm{m}=1 \mathrm{km}$ and $m_{\sigma}=0.05 \mathrm{km}$. The matter energy density at the center of the neutron star is chosen to be $\rho_{0}=0.001 \mathrm{km}^{-2}$.
}
\label{fig:ns1}
\end{figure}

From the ${}^r_t$ component of the modified Einstein equation, because of $G_{t}^{r}=T^r_t=T^{(\si)}{}_{t}^{r}=0$, we must have $X^r_t=0$. This can be solved by
\be
\label{k3sol}
k_{3}=\frac{\pm \sqrt{\alpha^{2}-\beta}-\alpha}{\beta}  ,
\ee
which generically have two branches of solutions, to be labelled as $k_{3+}$ and $k_{3-}$ respectively.  The ${}^t_r$ and ${}^r_r$ component of the Einstein equation reduce respectively to
\bal
\label{eq1}
\frac{a c^{\prime}}{r}-a c \sigma^{\prime 2}-8 \pi(p+\rho) c^{2}&=0  ,
\\
\label{eq2}
\frac{a-\left(1-k_{3}\right)^{2} c+r a^{\prime}}{r^{2}}-\frac{a \sigma^{\prime 2}}{2}&
\nn
-8 \pi p c +c W-k_{3}\left(2+\alpha k_{3}\right) c V&=0  ,
\eal
and the scalar equation of motion reduces to
\bal
\label{eq3}
\frac{2 c^{2} k_{3}\left(-2+k_{3}-\alpha k_{3}+2 \sqrt{c}+\alpha k_{3} \sqrt{c}\right) V_{\sigma}}{\sqrt{c}}&
\\
+\frac{4 a c \sigma^{\prime}}{r}+2 a c \sigma^{\prime \prime}+2 a^{\prime} c \sigma^{\prime}-a c^{\prime} \sigma^{\prime}-2 c^{2} W_{\sigma}&=0  .
\eal
Note that $e$ can be obtained from the ${}^\thi_\thi$ component of the modified Einstein equation. In the above three equations,  $a,~c,~\sigma,~p,~\rho$ are unknowns, thus more equations needed to close the system.  The two extra equations come from:  the matter energy-momentum conservation, which implies
\be
\label{eq4}
(p+\rho) a^{\prime}+2 a p^{\prime}=0  ,
\ee
and the knowledge of the equation of state  for the relativistic star
\be
\label{eq5}
p=p(\rho)   .
\ee
Thus, we can solve Eqs.~(\ref{eq1}, \ref{eq2}, \ref{eq3}, \ref{eq4}, \ref{eq5}) to obtain the compact star solution in mass-varying massive gravity.

Near the center of the star, we can taylor-expand
\bal
a(r)&=\sum_{n=0} a_{n} r^{n},~c(r)=\sum_{n=0} c_{n} r^{n},~\si(r)&=\sum_{n=0}  \si_{n} r^{n}  ,
\nn
p(r)&=\sum_{n=0}  p_{n} r^{n},~~\ri(r)=\sum_{n=0}  \ri_{n} r^{n}   .
\eal
Regularity of the solution at the center requires that $a_1=c_1=\si_1=0$. Plugging these into the equations of motion, we can perturbatively solve the equations of motion. To order $\mc{O}(r^2)$, we get
\bal
a(r)&=a_{0}+\frac{a_{0} r^{2}}{3\left(1- k_{3}\right)^{2}}\bigg[12 \pi  p_{0}+4 \pi  \rho_{0}-\frac{ m_{\sigma}^{2} \sigma_{0}^{2}}{2}
\nn
&~~~ + k_{3}\left(2+k_{3} \alpha\right)\left(m_{0}^{2}+m^{2} \sigma_{0}^{4}\right)\bigg]+\mc{O}\left(r^{3}\right)  ,
\\
c(r)&=\frac{a_{0}}{\left(1-k_{3}\right)^{2}}+\frac{4 a_{0} \pi\left(p_{0}+\rho_{0}\right)}{\left(1-k_{3}\right)^{4}} r^{2}+\mc{O}\left(r^{3}\right)  ,
\\
\sigma(r)&=\sigma_{0}+\frac{r^{2}}{6\left(1-k_{3}\right)^{2}}\bigg[ m_{\sigma}^{2} \sigma_{0}  -4 k_{3} m^{2} \sigma_{0}^{3}\left(k_{3} \alpha+2\right)
\nn
&- 4 k_{3} m^{2} \sigma_{0}^{3} \frac{\left|1-k_{3}\right|}{\sqrt{a_{0}}}\left(k_{3}-k_{3} \alpha-2\right)   \bigg] +\mc{O}\left(r^{3}\right)  ,
\\
p(r)&=p_{0}-\frac{\left(p_{0}+\rho_{0}\right)r^{2}}{6 \left(1-k_{3}\right)^{2}}\bigg[12  \pi p_0+4 \pi \rho_{0}-\frac{ m_{\sigma}^{2} \sigma_{0}^{2}}{2}
\nn
&~~~ + k_{3}\left(2+k_{3} \alpha\right)\left(m_{0}^{2}+m^{2} \sigma_{0}^{4}\right)\bigg] +\mc{O}\left(r^{3}\right)   ,
\\
\rho(r)&=\rho_{0}+\rho_{2}\left(\rho_{0}, \sigma_{0}\right) r^{2}+\mc{O}\left(r^{3}\right)  ,
\eal
where $p_0$ is determined by $\rho_0$ via the equation of state $p_{0}(\rho_0)$ and $\rho_2$ depends on $\rho_0$ and $\sigma_0$ and is easily determined once the functional form of the equation of state is specified. Therefore, the free parameters at the center are $\ri_0$, $a_0$ and $\si_0$. However, as we shall see, two of them will be fixed by the asymptotical flatness at infinity, up to some discrete sets of values for, say, $(a_0, \sigma_0)$.

Numerically, we integrate Eqs.~(\ref{eq1}, \ref{eq2}, \ref{eq3}, \ref{eq4}, \ref{eq5}) from a point near the center of the star ($r/{\rm km}=10^{-8}$) to the boundary of the star $r=r_s$ where $\ri(r_s)\to 0$ (the threshold being set at $10^{-10}{\rm km}^{-2}$), and then set $\ri(r)=p(r)=0$ and integrate from the boundary to a large $r$. Via a shooting procedure, we can tune $(a_0, \sigma_0)$ to obtain asymptotically flat solutions such that $\sigma(r\to \infty)\to0$ and $b(r\to \infty)\to 0$. Our model Eq.~(\ref{VWform}) is $Z_2$ symmetric in $\si$, so without lost of generality we can choose the value of $\si$ to be positive near the center.

By Eq .~(\ref{k3equ}), we can infer that $d=r^2/(1-k_3)^2$, which means that $r$ is not the radial coordinate that asymptotes to the standard Minkowski metric $\eta_{\mu\nu}$ at infinity. The standard asymptotically Minkowski radial coordinate can be obtained by a constant re-scaling
\be
R=\f{r}{|1-k_3|} .
\ee
such that in the new radial coordinate we have
\be
\label{dsrescale}
\mathrm{d} s^{2}=-A(R) \mathrm{d}t^{2}+2 B(R) \mathrm{d}R \mathrm{d}t+E(R) \mathrm{d}R^{2}+R^2 \mathrm{d} \Omega^{2}   ,
\ee
where $A(R)=a(r)$, $B(R)=|1-k_3|b(r)$ and $E(R)=e(r)(1-k_3)^2$. We will present the numerical results using this metric.

\section{Neutron stars}
\label{sec:ns}

As mentioned in the last section, to obtain a neutron star solution, we need the input of the equation of state $p(\rho)$. However, the exotic nature of matter inside the neutron star, particularly in the strong gravity environment, is far from what we know or can probe in the current particle/nuclear physics. Therefore, our knowledge about the equation of state of a neutron star is rather limited. There are quite a few theoretical models for the neutron star equation of state such as the APR model \cite{Akmal:1998cf}, the SLy model \cite{Douchin:2001sv}, the Shen Model \cite{Shen:1998by}, etc.

\begin{figure}
\centering
\includegraphics[width=0.4\textwidth]{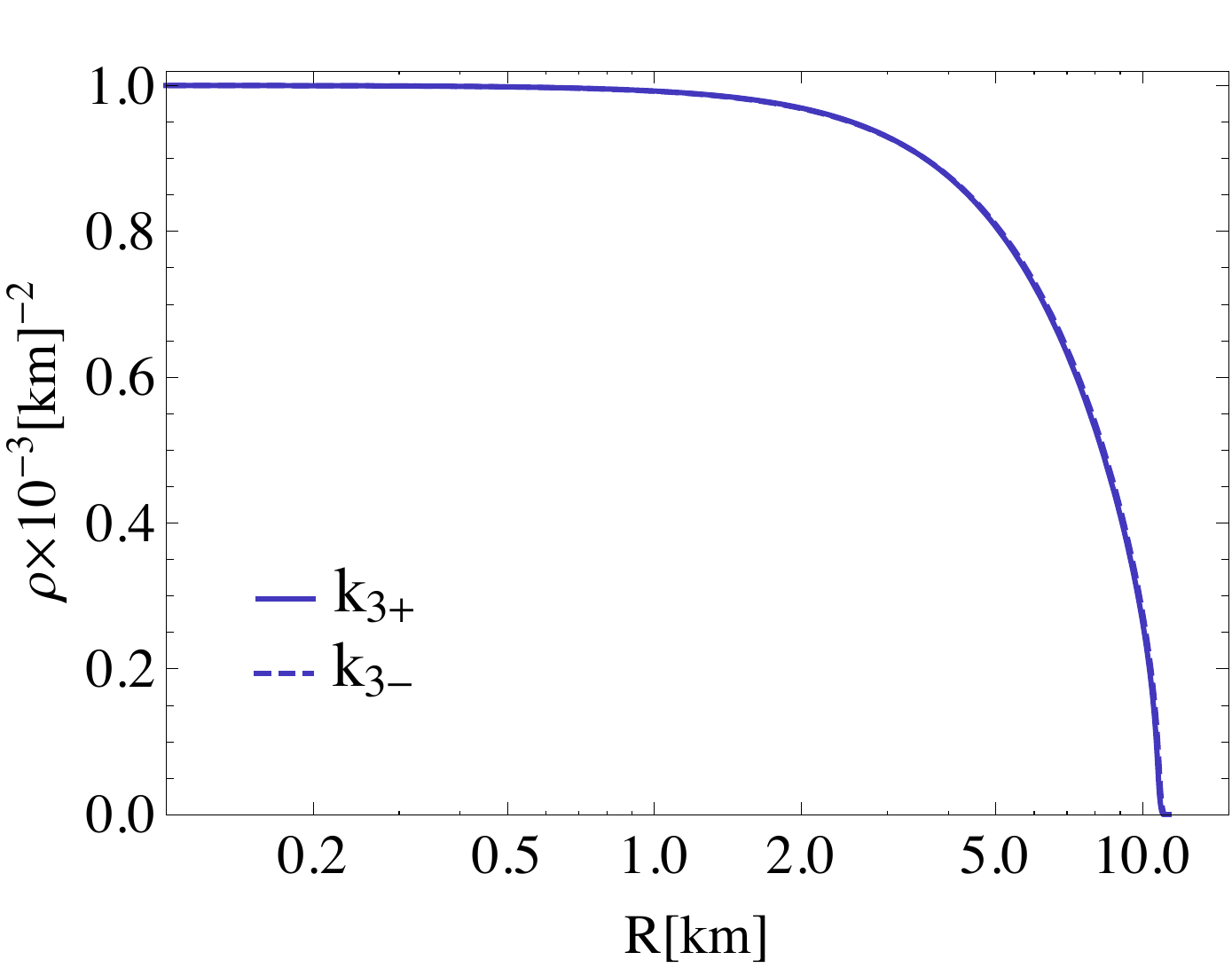}
\includegraphics[width=0.4\textwidth]{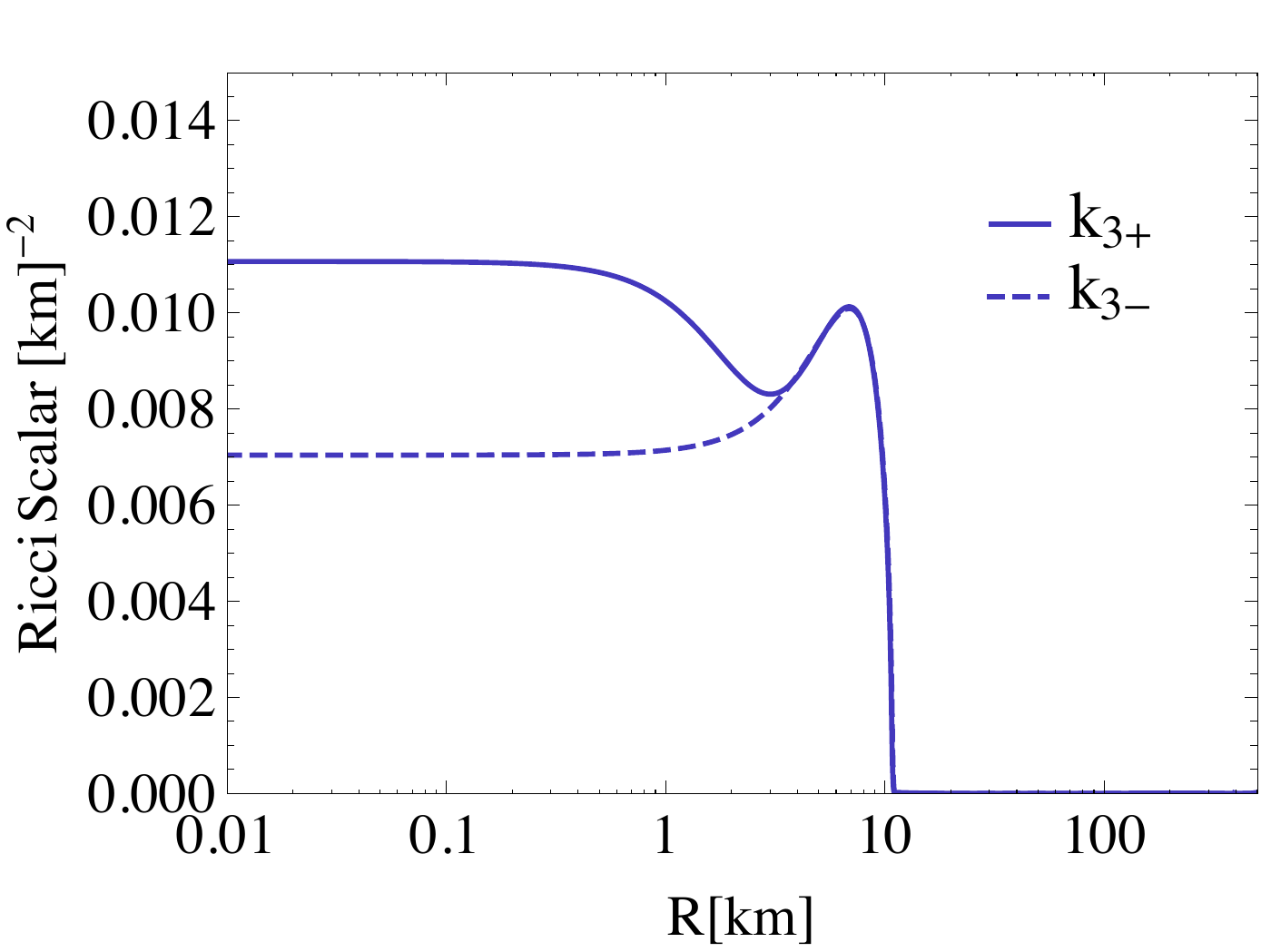}
\caption{Matter energy density $\rho$ and Ricci scalar curvature for the Neutron star solutions in Fig.~\ref{fig:ns1}. The radii of the stars for the ${k}_{3+}$ and ${k}_{3-}$ branches are $11.26$km and $11.30$km respectively.}
\label{fig:ns2}
\end{figure}

 In this paper, we will make use of the APR and SLy equation of state, both of which allow for a neutron star whose maximum mass is greater than the most massive neutron stars observed so far such as PSR J1614-2230 ($1.97\pm{0.04}M_\odot$) \cite{Demorest:2010bx} and PSR J0348+0432 ($2.01\pm{0.04}M_\odot$) \cite{Antoniadis:2013pzd}, $M_\odot$ being the solar mass. The APR equation of state mainly describes the liquid core of a neutron star. It is obtained by the variational chain summation method, using the new Argonne $\nu18$ two-nucleon interaction. When the two-nucleon boost corrections and three-nucleon interactions are also included, the maximum mass of the static neutron star can be increased to $2.20\ M_\odot$ \cite{Akmal:1998cf}. The SLy model, on the other hand, takes into account both the liquid core and the crust of the neutron star using the Skyrme-type effective nucleon-nucleon interaction, which provides a maximum mass limit of $2.05\ M_\odot$ for a static neutron star \cite{Douchin:2001sv}. The equation of state of the crust is obtained in the zero temperature approximation, while the equation of state of the liquid core is calculated when minimal $npe\mu$ composition is satisfied. Analytical fits of these equations of state have been obtained, which are easier to use for our purposes. Defining $\xi=\log \left(\rho / \mathrm{g} \operatorname{cm}^{-3}\right),~\zeta=\log \left(\mathrm{p} / \mathrm{g cm^{-1} s^{-2}}\right),~ f_{0}(x)={1}/{(e^{x}+1)}$, the equation of state can be written as \cite{Haensel:2004nu}
\bal
\zeta &=\frac{a_{1}+a_{2} \xi+a_{3} \xi^{3}}{1+a_{4} \xi} f_{0}\left(a_{5}\left(\xi-a_{6}\right)\right)
\nn
&~~~+\left(a_{7}+a_{8} \xi\right) f_{0}\left(a_{9}\left(a_{10}-\xi\right)\right)
\nn
&~~~+\left(a_{11}+a_{12} \xi\right) f_{0}\left(a_{13}\left(a_{14}-\xi\right)\right)
\nn
&~~~+\left(a_{15}+a_{16} \xi\right) f_{0}\left(a_{17}\left(a_{18}-\xi\right)\right)   ,
\eal
where the fitting parameters for the two models are given in Table \ref{table:eos}.

\begin{figure}
\centering
\includegraphics[width=0.4\textwidth]{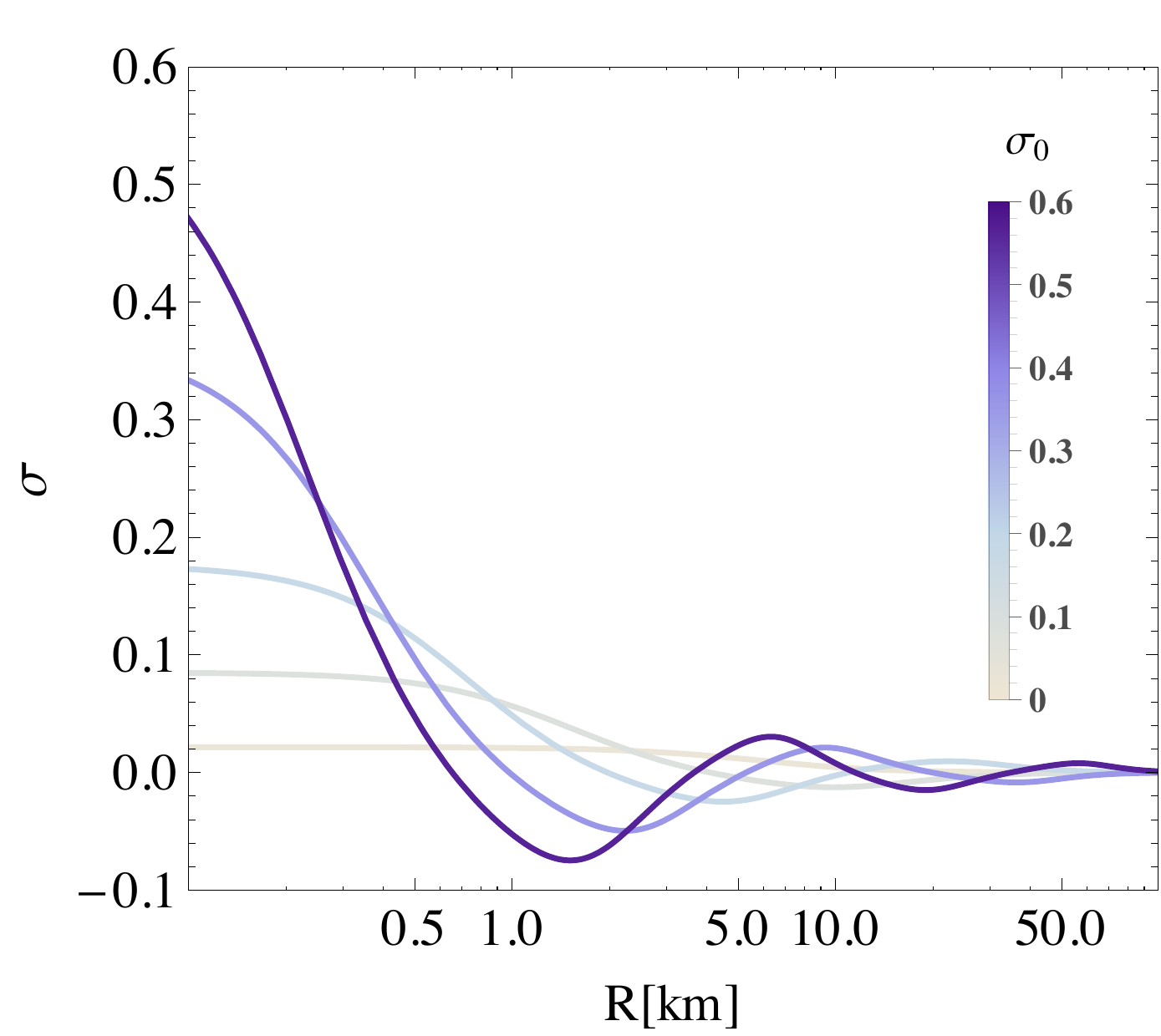}
\includegraphics[width=0.4\textwidth]{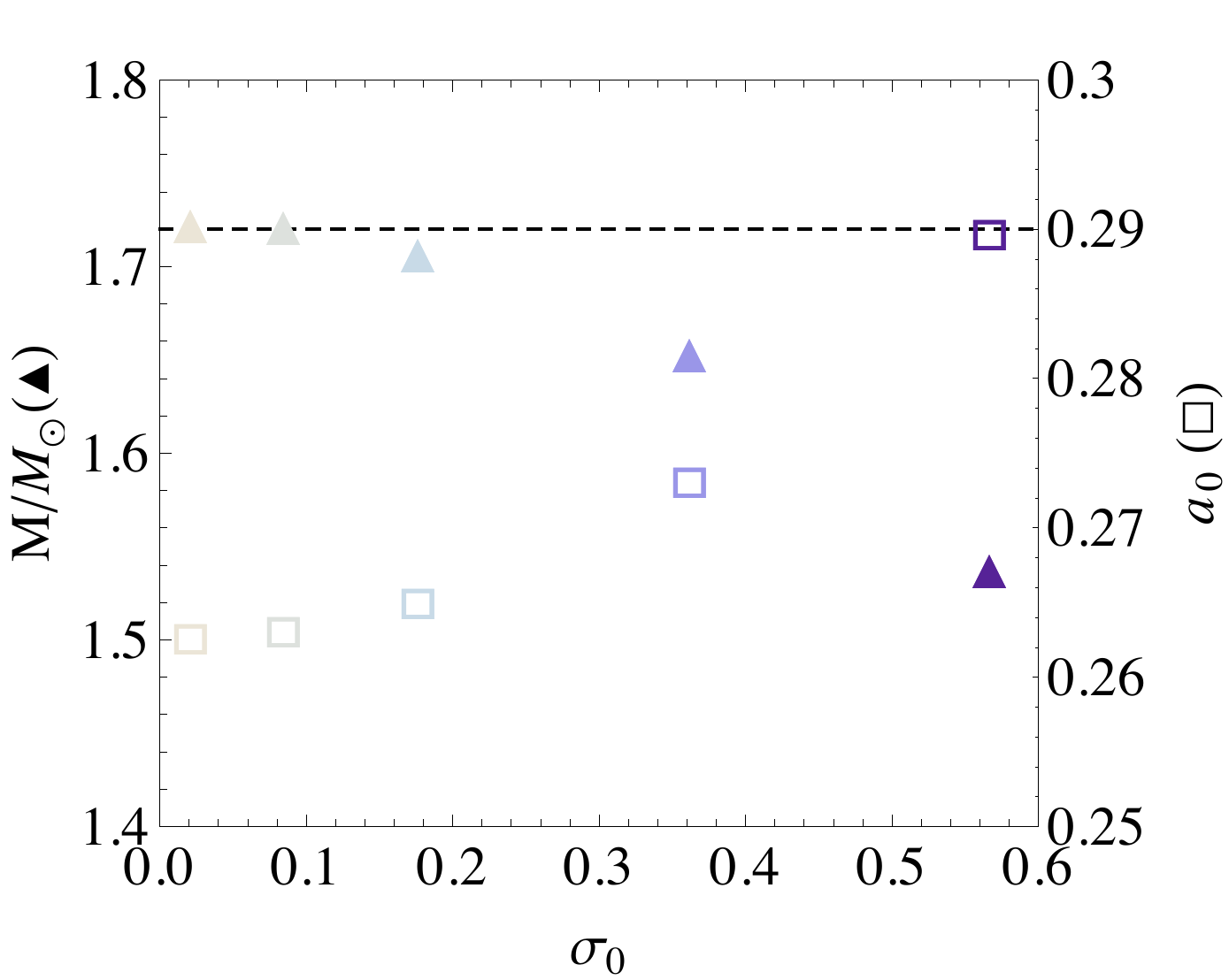}
\caption{A sequence of neutron star solutions where the $\sigma$ field ``oscillates'' a number of times before setting down to zero at large $R$. The bottom plot depicts the parameters of the solutions in the top plot. The graviton potential parameters are $\alpha=2, \beta=3$ and the ${k}_{3+}=-1 / 3$ branch is chosen. The mass parameters are chosen as ${m}=10 \mathrm{km}$ and $m_{\sigma}=0.05 \mathrm{km}$.  In the bottom plot, the left vertical axis labels the ADM mass and the right vertical axis labels the $a_0$ parameter. The horizontal dashed line corresponds to the ADM mass in general relativity.}
\label{fig:nsS}
\end{figure}

\begin{table}[ht]
\caption{Equation of state -- neutron star} 
\centering 
\begin{tabular}{c c c c c c} 
\hline\hline 
$a_i$ & SLy & APR & $a_i$ & SLy & APR\\ 
\hline 
$a_1$ & 6.22 \;& 6.22 \;& $a_{10}$ & 11.4950 \;& 11.5756 \\
$a_2$ & 6.121 \;& 6.121 \;& $a_{11}$ & -22.775 \;& -42.489 \\
$a_3$ & 0.005925 \;& 0.006035 \;& $a_{12}$ & 1.5707 \;& 3.8175 \\
$a_4$ & 0.16326 \;& 0.16354 \;& $a_{13}$ & 4.3 \;& 2.3 \\
$a_5$ & 6.48 \;& 4.73 \;& $a_{14}$ & 14.08 \;& 14.81 \\
$a_6$ & 11.4971 \;& 11.5831 \;& $a_{15}$ & 27.80 \;& 29.80 \\
$a_7$ & 19.105 \;& 12.589 \;& $a_{16}$ & -1.653 \;& -2.976 \\
$a_8$ & 0.8938 \;& 1.4365 \;& $a_{17}$ & 1.50 \;& 1.99 \\
$a_9$ & 6.54 \;& 4.75 \;& $a_{18}$ & 14.67 \;& 14.93 \\[1ex] 
\hline 
\end{tabular}
\label{table:eos} 
\end{table}

The radial profiles of representative neutron star solutions are given in Fig.~\ref{fig:ns1} and Fig.~\ref{fig:ns2}. As we can see, for a given set of parameters, there are two branches of solutions, corresponding to the different sign choices in Eq.~(\ref{k3sol}). Although very close to each other, the metric quantities $A$ and $C$ and the matter energy density $\rho$ for the two branches are not the same but differ slightly from each other, while the scalar field $\si$ is relatively different for the two branches. The Ricci scalars of the two branches differ significantly well inside the star but are almost the same when approaching to the surface and outside the star.

\begin{figure}
\centering
\includegraphics[width=0.4\textwidth]{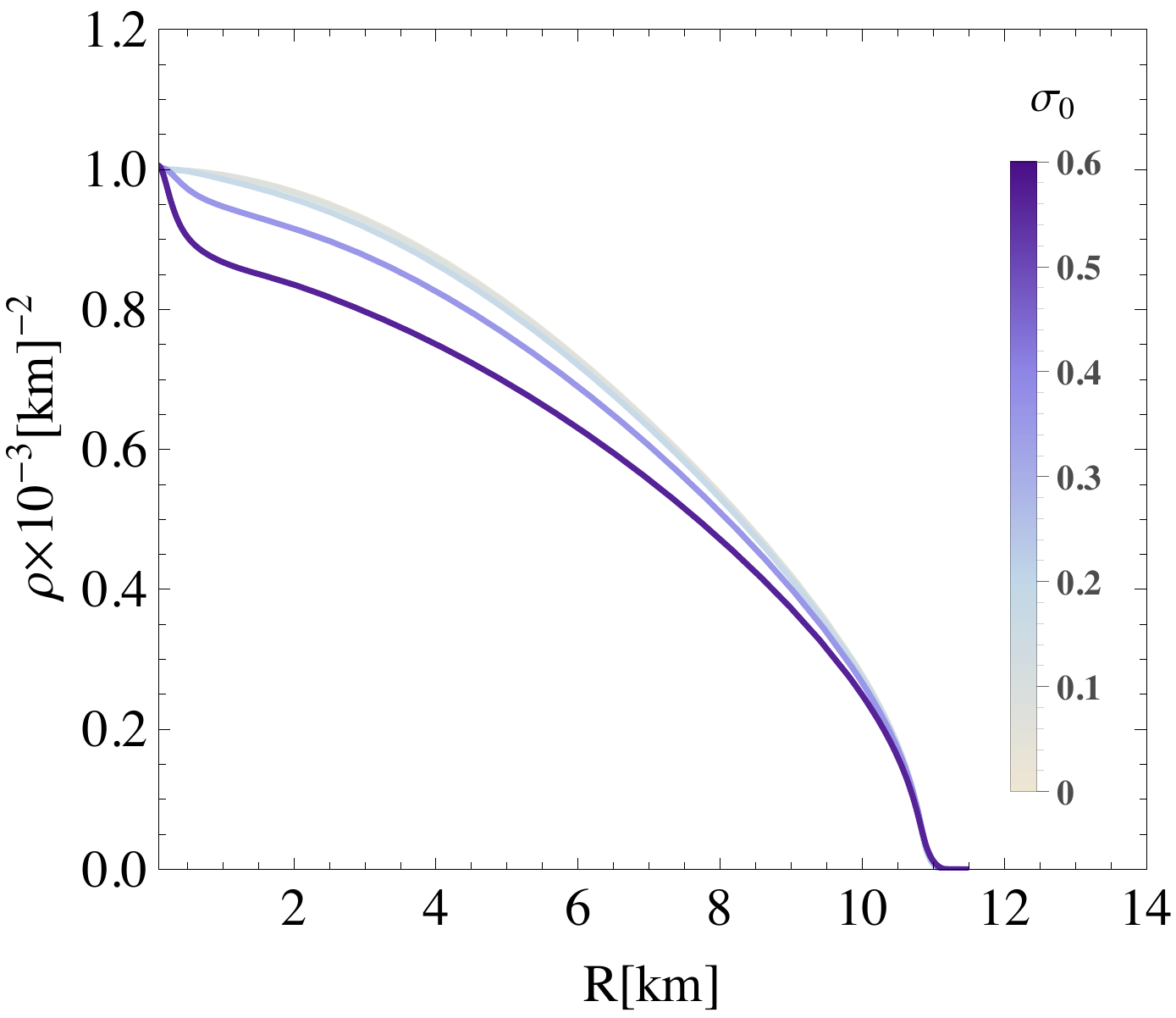}
\caption{Matter energy densities for the different neutron star solutions in Fig.~\ref{fig:nsS}. We see that the mass of the neutron star $M$ becomes smaller for a larger $\sigma_0$ because the matter energy density decreases when $\sigma_0$ increases. }
\label{fig:nsrhosi}
\end{figure}

\begin{figure}
\centering
\includegraphics[width=0.4\textwidth]{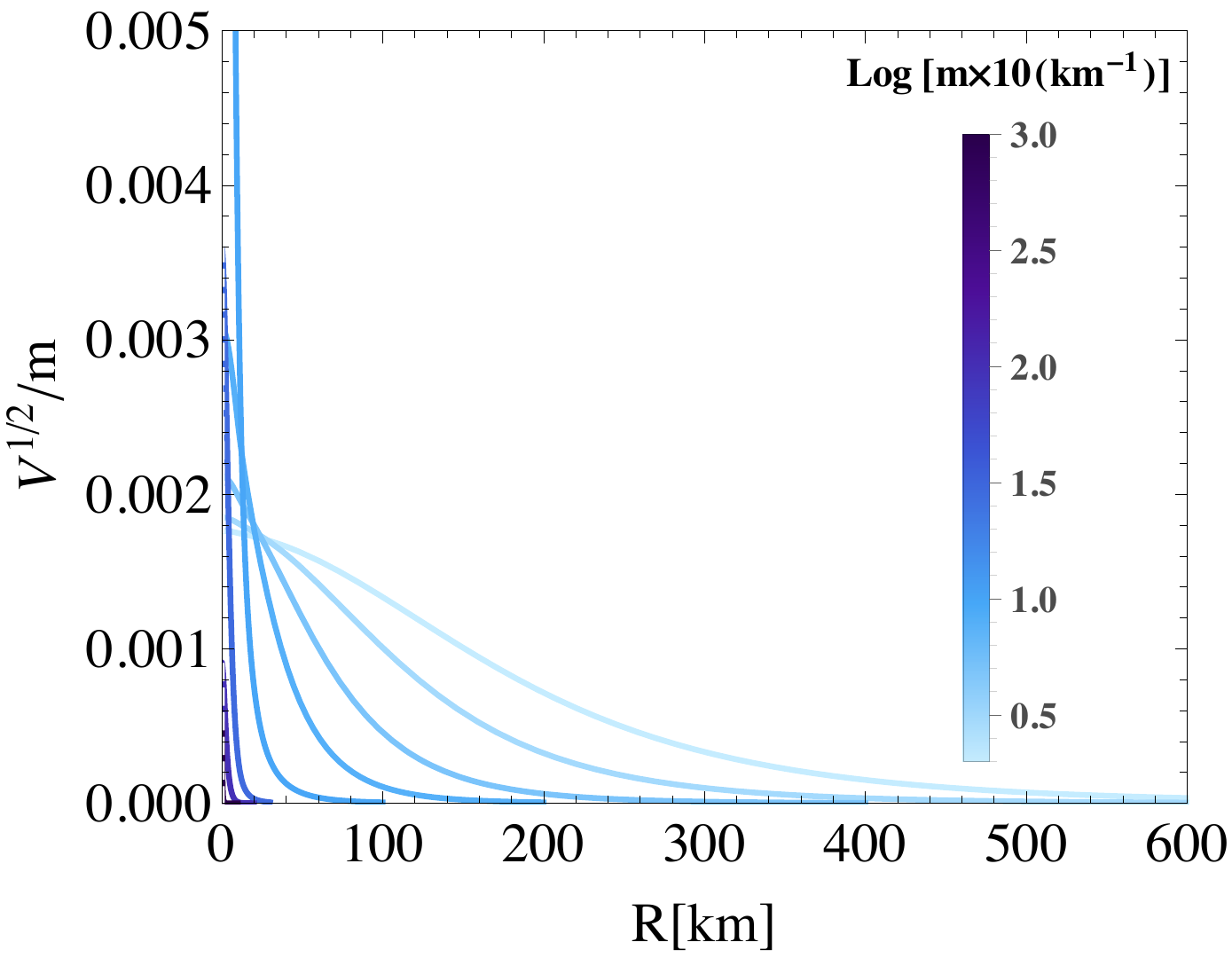}
\caption{Radial profile of the effective graviton mass $V^{1/2}$. The graviton potential parameters are $\alpha=2, \beta=3$ and the $\mathrm{k}_{3+}=-1 / 3$ branch is chosen. We fix $m_\sigma=m/100$.  The logarithm is with base 10.  The effective graviton mass becomes sizable near the center of the star and decreases to negligible values away from the center.}
\label{fig:nsV}
\end{figure}

For the solutions in Fig.~\ref{fig:ns1}, the scalar field quickly falls off to zero radially. But this is not the only solution we can have. By tuning different values of $(\si_0, a_0)$, we can actually find a sequence of asymptotically flat solutions for a given $\rho_0$. For these solutions, the $\si$ field ``oscillates'' a number of times before setting down to zero at infinity (see Fig.~\ref{fig:nsS}). For ${m}=10 \mathrm{km}$ and $m_{\sigma}=0.05 \mathrm{km}$, we can find four extra such kind of neutron star solutions, and for each extra half cycle of oscillation (up to a certain number of oscillations) there is a unique extra solution. But the number of the extra solutions seems to depend on the values of $m$ and $m_\si$, at least under our current numerical accuracy. In particular, we find that the number of solutions increases for bigger $m$ and as well as for smaller $m_\si$.

One intriguing observation of the bottom plot of Fig.~\ref{fig:nsS} is that there is some kind of approximate reflection symmetry between the ADM mass $M$ at infinity and $a_0$ near the center of the star. Note that in our numerical scheme we integrate from the center of the star to the infinity for which we choose appropriate $\si_0$ and $a_0$ to find a solution; Alternatively, we could have integrated from the infinity to the center for which we choose appropriate $\si_0$ and $M$ to find the same solution. Therefore, this apparent reflection symmetry suggests that there is some kind of symmetric connection between the two sets of boundary data. Also, we see that for bigger $\si_0$ the ADM mass can be significantly smaller than the corresponding value in general relativity. This is because when $\si_0$ increases the matter energy density $\rho$ decreases, most significantly near the center (see Fig.~\ref{fig:nsrhosi}), which seems to suggest that for larger $\si_0$ the scalar forces can significantly alleviate the gravitational attraction.

Similar to the case of hairy black holes \cite{Zhang:2017jze}, the effective graviton mass in this model can become very large near the center of the neutron star but decrease very rapidly away from the neutron star (see Fig.~\ref{fig:nsV}), thus easily evading all the current constraints on the graviton mass \cite{deRham:2016nuf} and other tests of gravity \cite{Will:2014kxa}.

\begin{figure}
\centering
\includegraphics[width=0.4\textwidth]{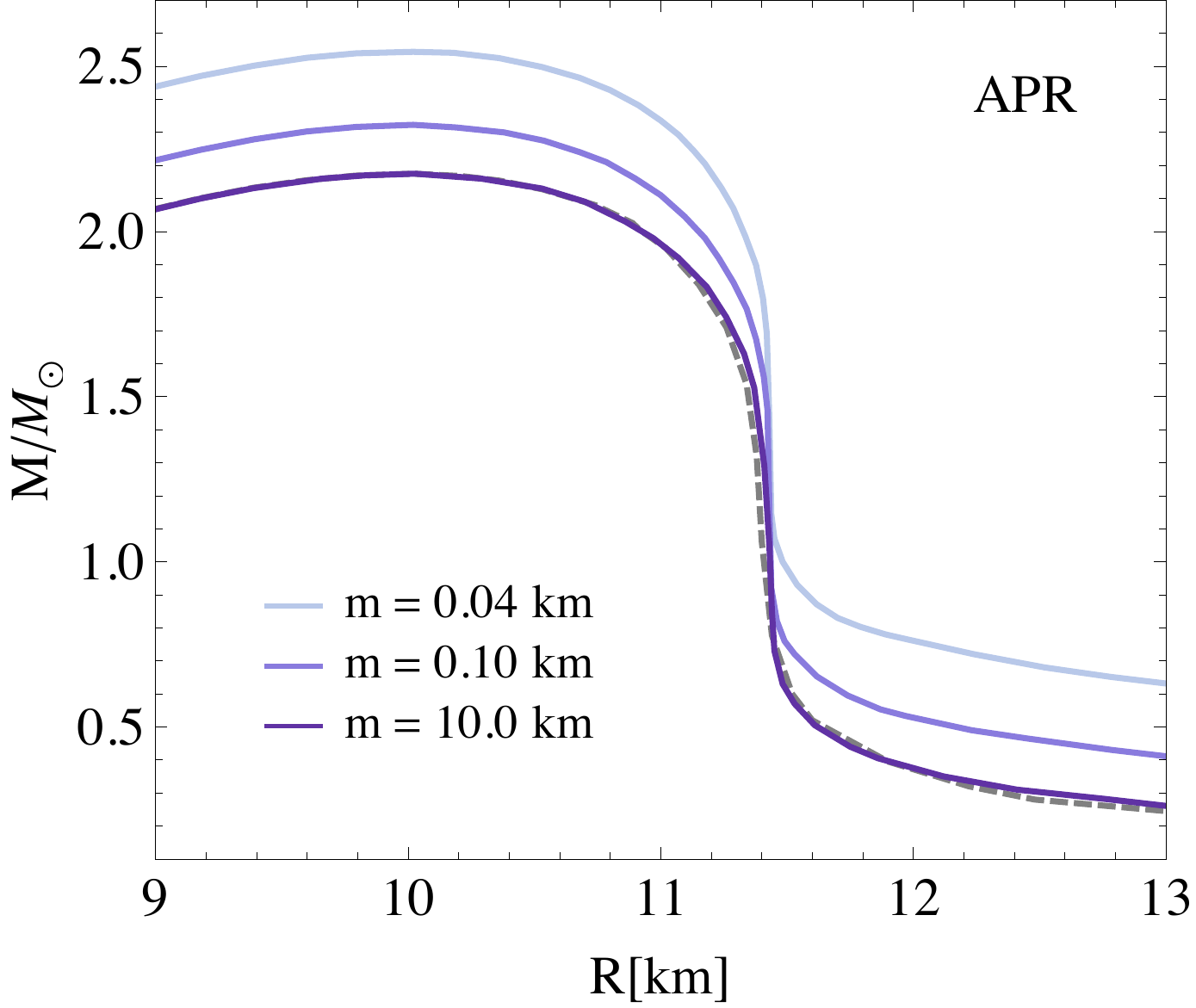}
\includegraphics[width=0.4\textwidth]{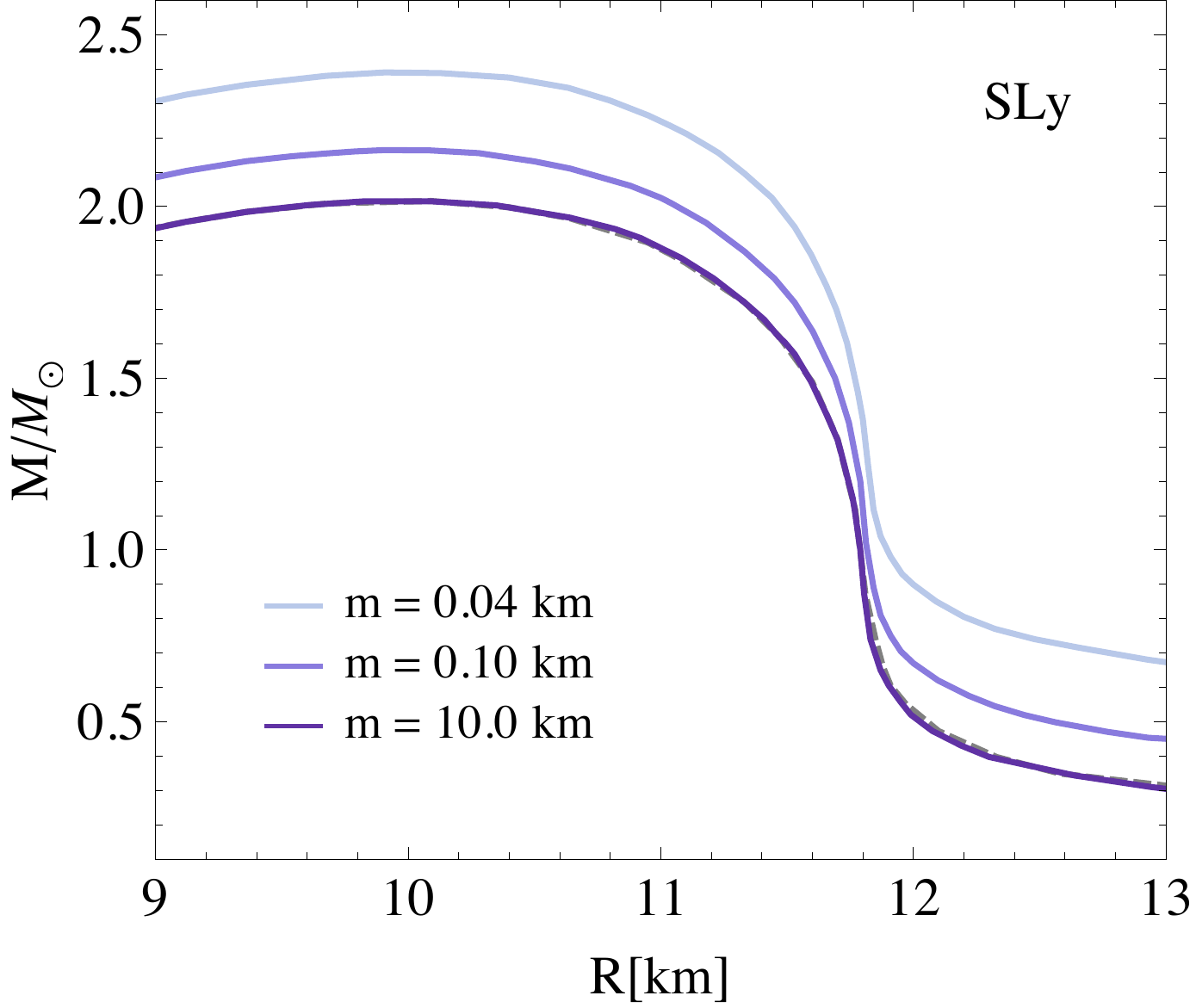}
\caption{Mass-radius relations for neutron stars for different $m$. The top plot is for the APR model and the bottom one is for the SLy model. The graviton potential parameters are $\alpha=2, \beta=3$ and the $\mathrm{k}_{3+}=-1 / 3$ branch is chosen. We fix $m_\sigma=m/100$. }
\label{fig:nsMRm}
\end{figure}

\begin{figure}
\centering
\includegraphics[width=0.4\textwidth]{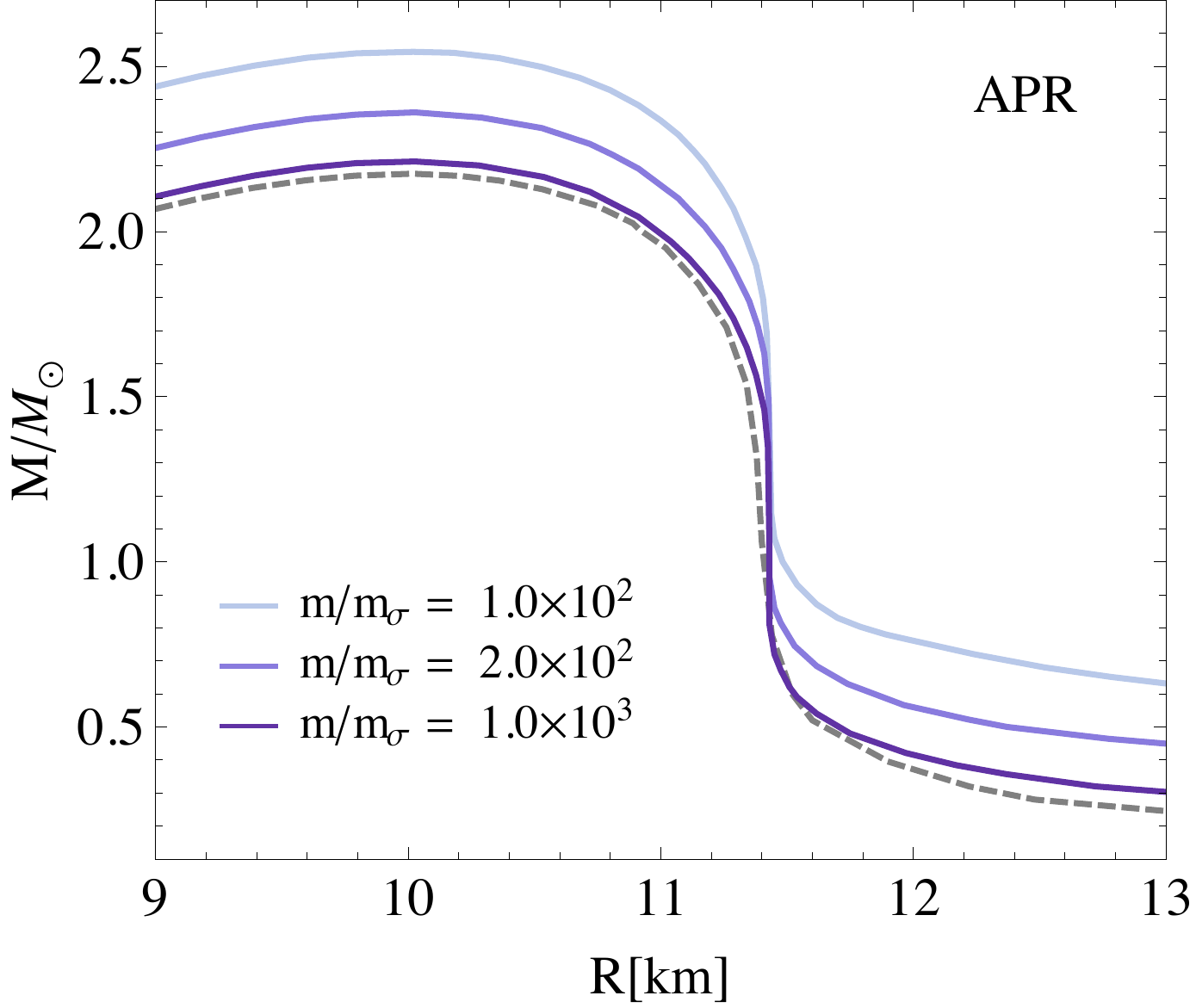}
\includegraphics[width=0.4\textwidth]{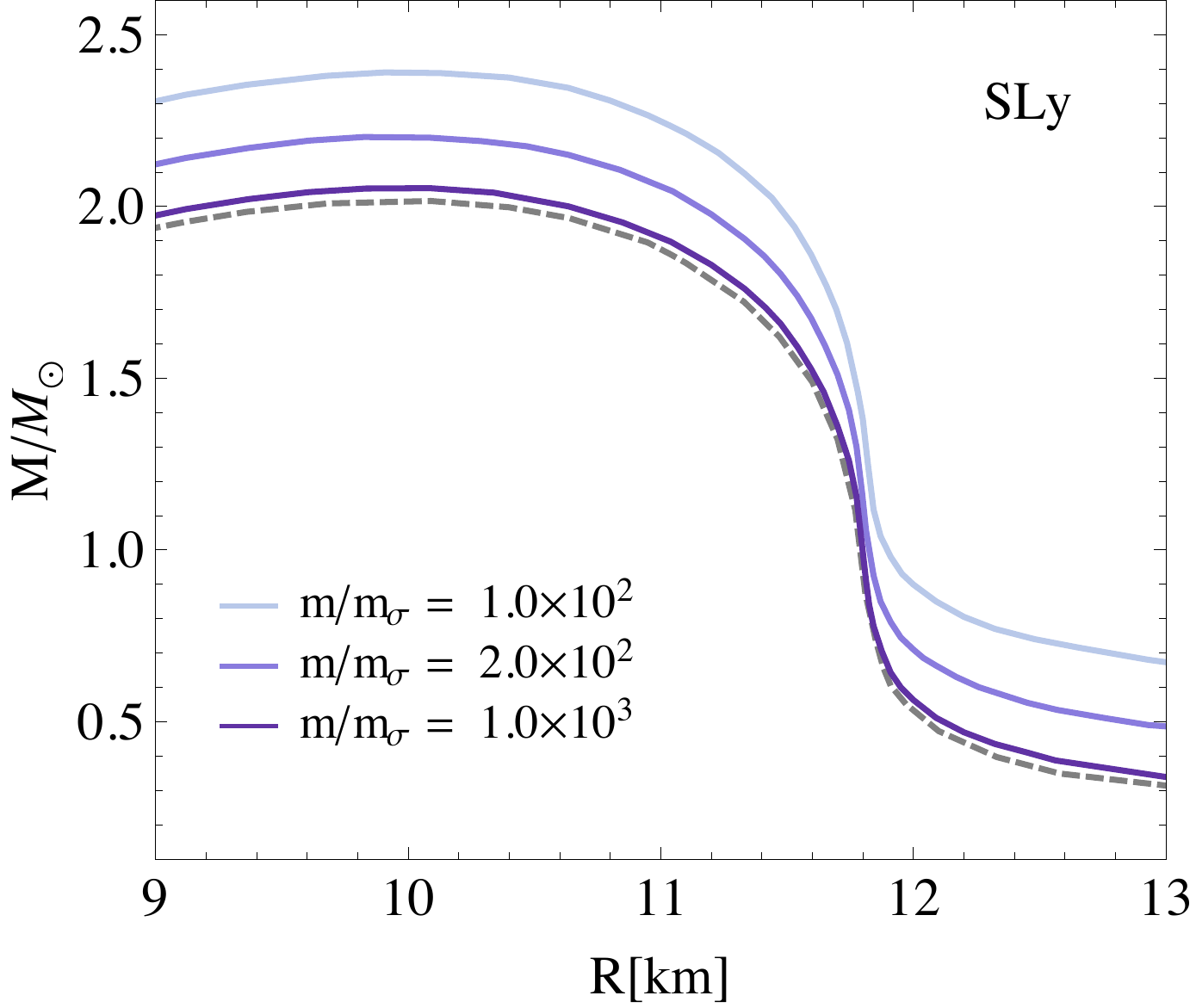}
\caption{Mass-radius relations for neutron stars for different $m$. The top plot is for the APR model and the bottom one is for the SLy model. The graviton potential parameters are $\alpha=2, \beta=3$ and the $\mathrm{k}_{3+}=-1 / 3$ branch is chosen. We fix $m=0.04{\rm km}$. }
\label{fig:nsMRmmsi}
\end{figure}

We have computed the mass-radius relation of the neutron star for both the APR and SLy models for different parameters in MVMG; see Fig.~\ref{fig:nsMRm} and Fig.~\ref{fig:nsMRmmsi}. We see that in mass-varying massive gravity, for the same equation of state, the maximum mass of the neutron star can be raised for both the APR and SLy model, compared to that in general relativity. The patterns of the modification to the mass-radius relation are quite similar for the APR and SLy model. Interestingly, for the steepest part of the mass-radius relation, the modifications are small for both of these two models.

\section{White dwarfs}
\label{sec:wd}

\begin{figure}
\centering
\includegraphics[width=0.4\textwidth]{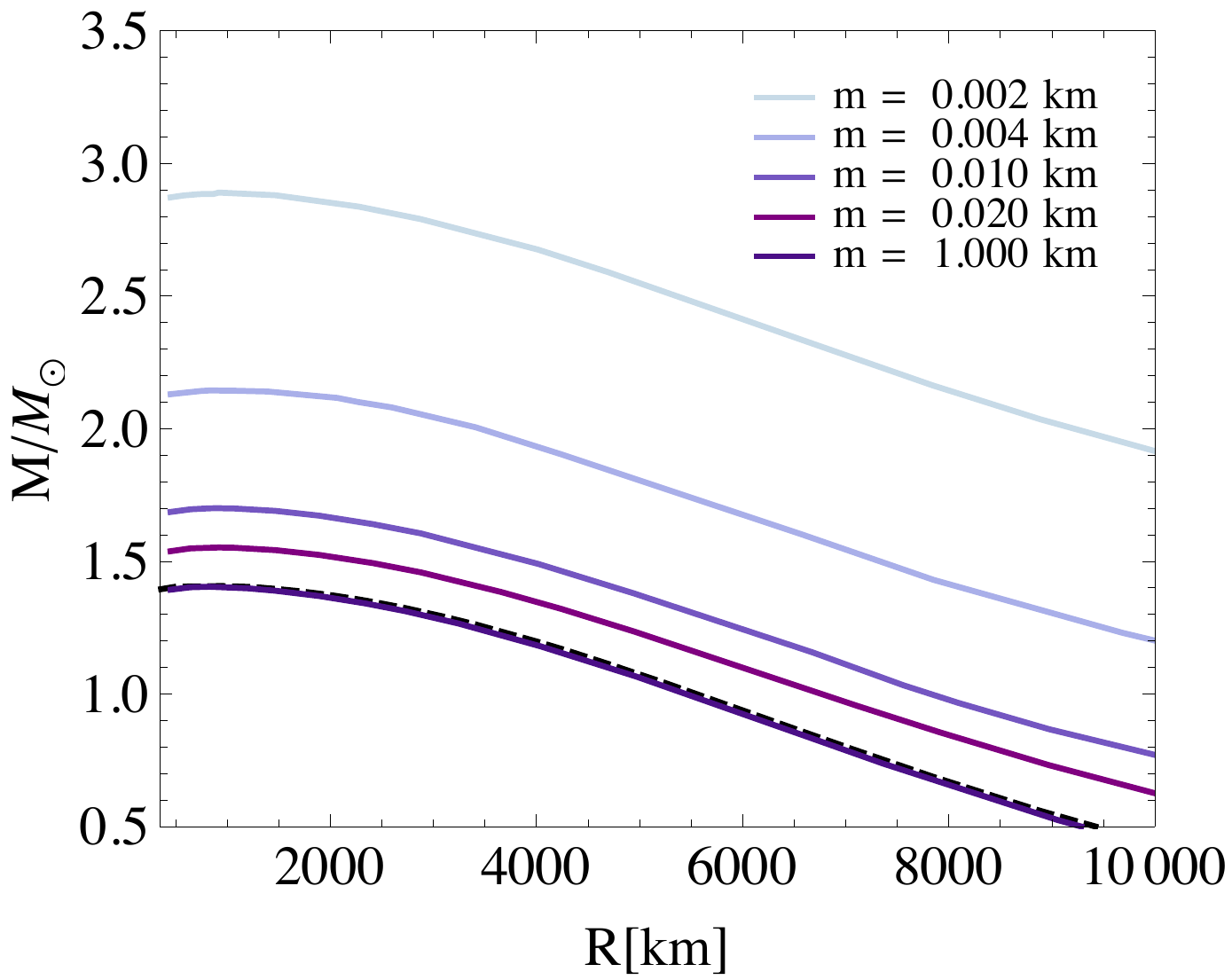}
\includegraphics[width=0.4\textwidth]{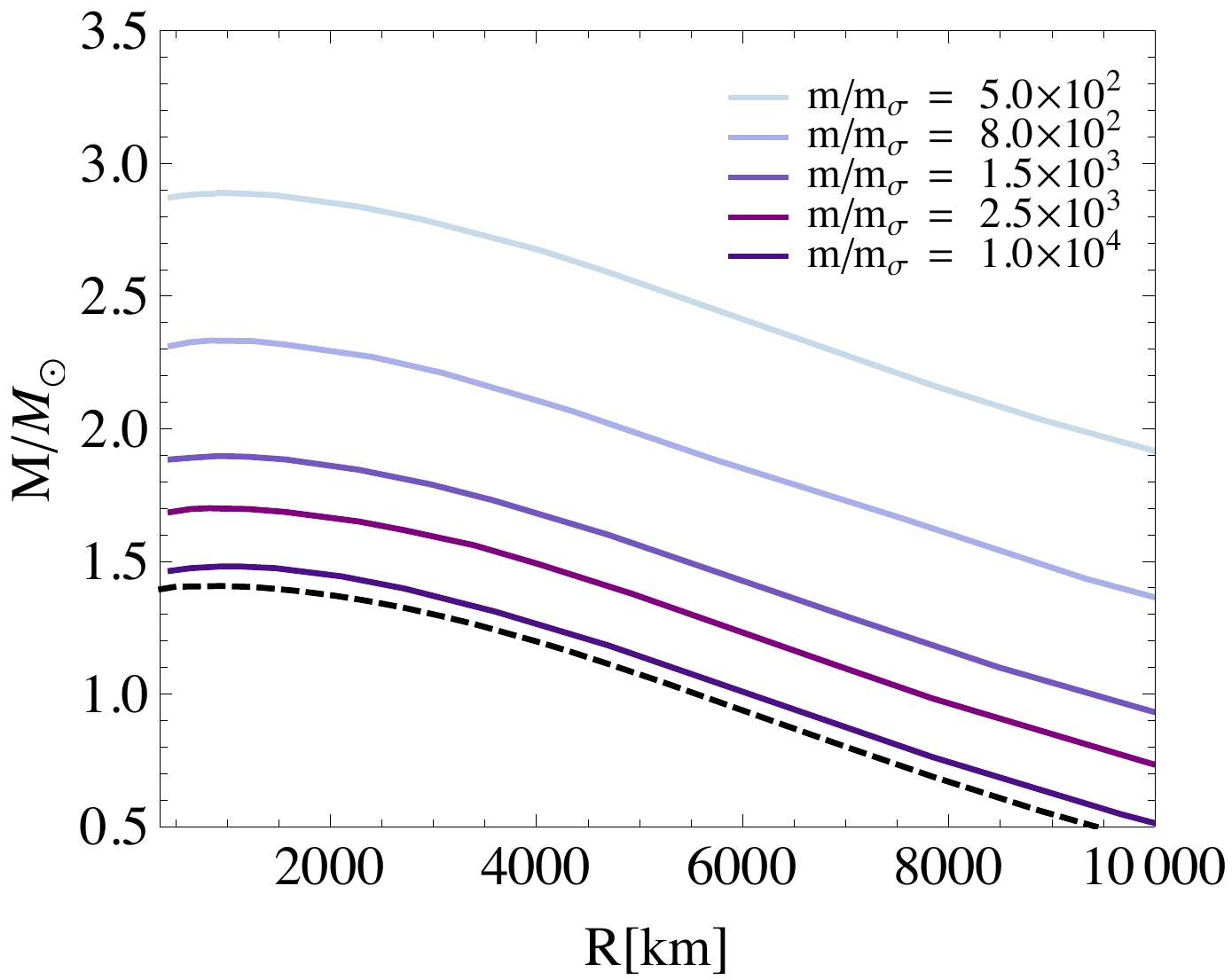}
\caption{Mass-radius relations for white dwarfs. The graviton potential parameters are $\alpha=2, \beta=3$ and the $\mathrm{k}_{3+}=-1 / 3$ branch is chosen. For the top plot we choose $m_\si/m=0.002$ and for the bottom plot we choose $m=0.002{\rm km}$.   }
\label{fig:wdMR}
\end{figure}

Now, we turn our attention to the white dwarf solutions. The procedure to obtain white dwarf solutions is mostly the same as the neutron stars, except, of course, we need to use the equation of state of white dwarfs, which are less dense than neutron stars.

For the white dwarfs, we employ the simple Chandrasekhar equation of state, which is given by \cite{Chandrasekhar:1931ih, Chandrasekhar:1935zz}
\bal
p&=\frac{ m_e^{4}}{24\pi^2\hbar^3 }\left[x\left(2 x^{2}-3\right)\left(x^{2}+1\right)^{\f12}+3 \sinh ^{-1} x\right]   ,
\eal
where $x=\sqrt[3]{3\pi^2 \rho \hbar^3/ (m_e^{3}m_p \mu_e )}$, $m_e$ is the electron mass, $m_p$ is the proton mass and we choose the mean molecular weight per electron $\mu_e=2$. Note that we have set the speed of light to 1. The classic Chandrasekhar limit states the mass of a white dwarf does not exceed 1.44 $M_\odot$ in Newtonian physics and 1.405 $M_\odot$ in general relativity. However, recent observations of type Ia supernovae seem to prefer a larger maximum mass for the white dwarf, and it has been suggested that there may exist white dwarfs with the mass ranging from $2.1M_\odot$ to $2.8M_\odot$ \cite{Howell:2006vn, Scalzo:2010xd, Hicken:2007ap, Yamanaka:2009dp, Silverman:2010bh, Taubenberger:2010qv}. To reconcile the theoretical predictions with observations, an improved equation of state may be considered. Here, alternatively, we see that the same can be achieved by replacing general relativity with a massive gravity theory. In Fig.~\ref{fig:wdMR}, we plot the mass-radius diagram for different model parameters, and find that a white dwarf with mass $2.9M_\odot$ can be obtained. Generically, a greater maximum white dwarf mass can be achieved for a smaller $m$ or smaller $m_\si$.

Probably not surprisingly, we also find sequences of white dwarf solutions where the scalar field oscillates radially to zero for a given central matter energy density, in addition to the solution where the scalar field decreases to zero directly, which is very much analogous to the case of neutron stars in Fig.~\ref{fig:nsS}.

\section{Comparison of different constraints}

\begin{figure}
\centering
\includegraphics[width=0.4\textwidth]{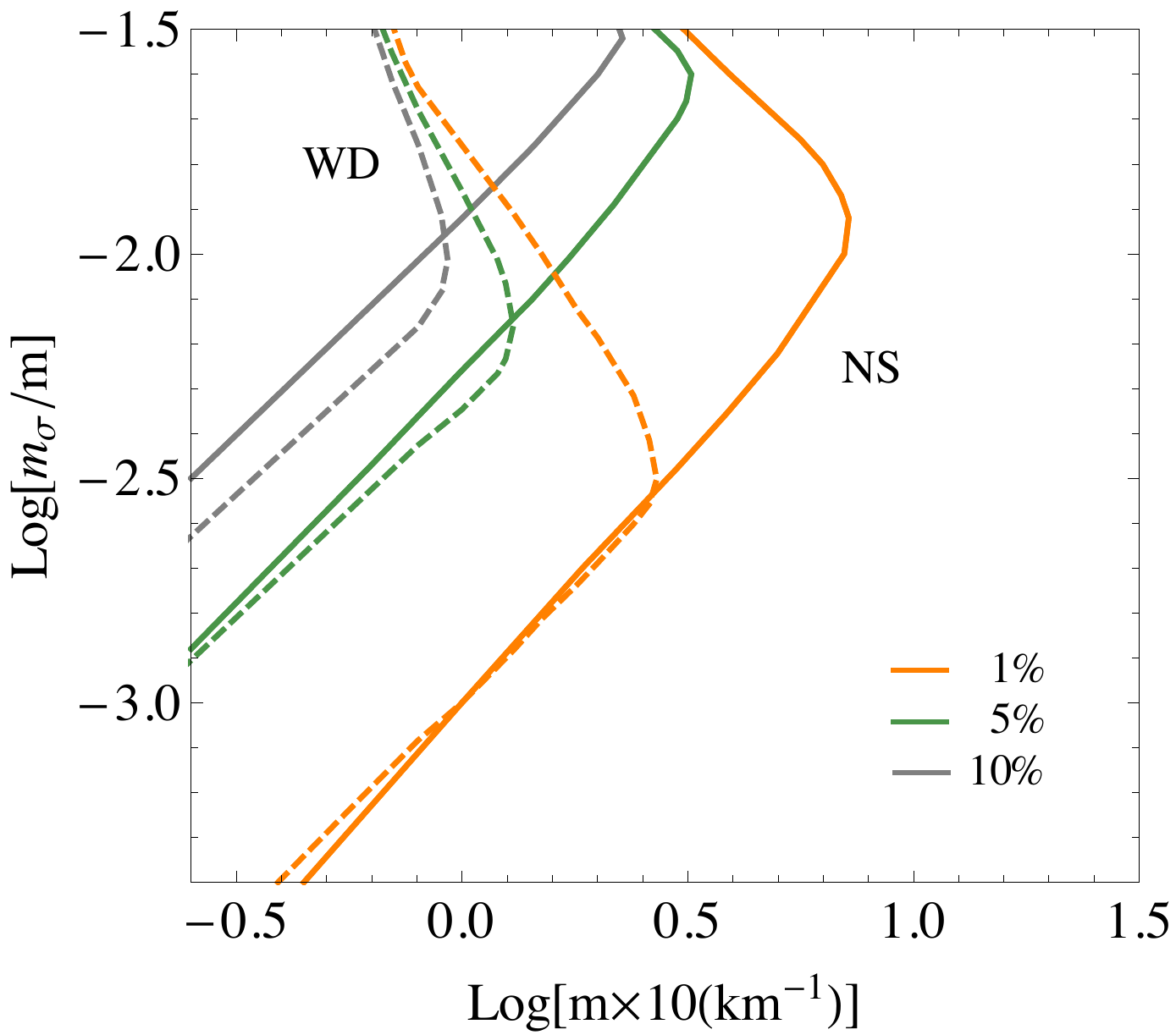}
\caption{Contours of constant ADM mass deviations from general relativity. The solid lines are for neutron stars and the dashed lines are for white dwarfs. For example, for every point in the solid orange line, the ADM mass of the star in MVMG deviates from that of general relativity by 1\%. The graviton potential parameters are $\alpha=2, \beta=3$ and the $\mathrm{k}_{3+}=-1 / 3$ branch is chosen.}
\label{fig:nswd}
\end{figure}

We may also compare the relative strengths between neutron stars and white dwarfs in constraining the MVMG parameter space in terms of the ADM mass deviations. In Fig.~\ref{fig:nswd}, we plot the contours of constant ADM mass deviations from general relativity for a neutron star with central matter energy density $\rho_0=0.001{\rm km}^{-2}$ and for a white dwarf with central matter energy density $\rho_0=1.2\times 10^{-8}{\rm km}^{-2}$. We see that greater $m_\si$ and smaller $m$ tend to have greater ADM mass deviations, and neutron stars tend to better constrain MVMG for larger $m$ and $m_\si$ while white dwarfs have a slight advantage for smaller $m$ and larger $m_\si$.

\begin{figure}
\centering
\includegraphics[width=0.4\textwidth]{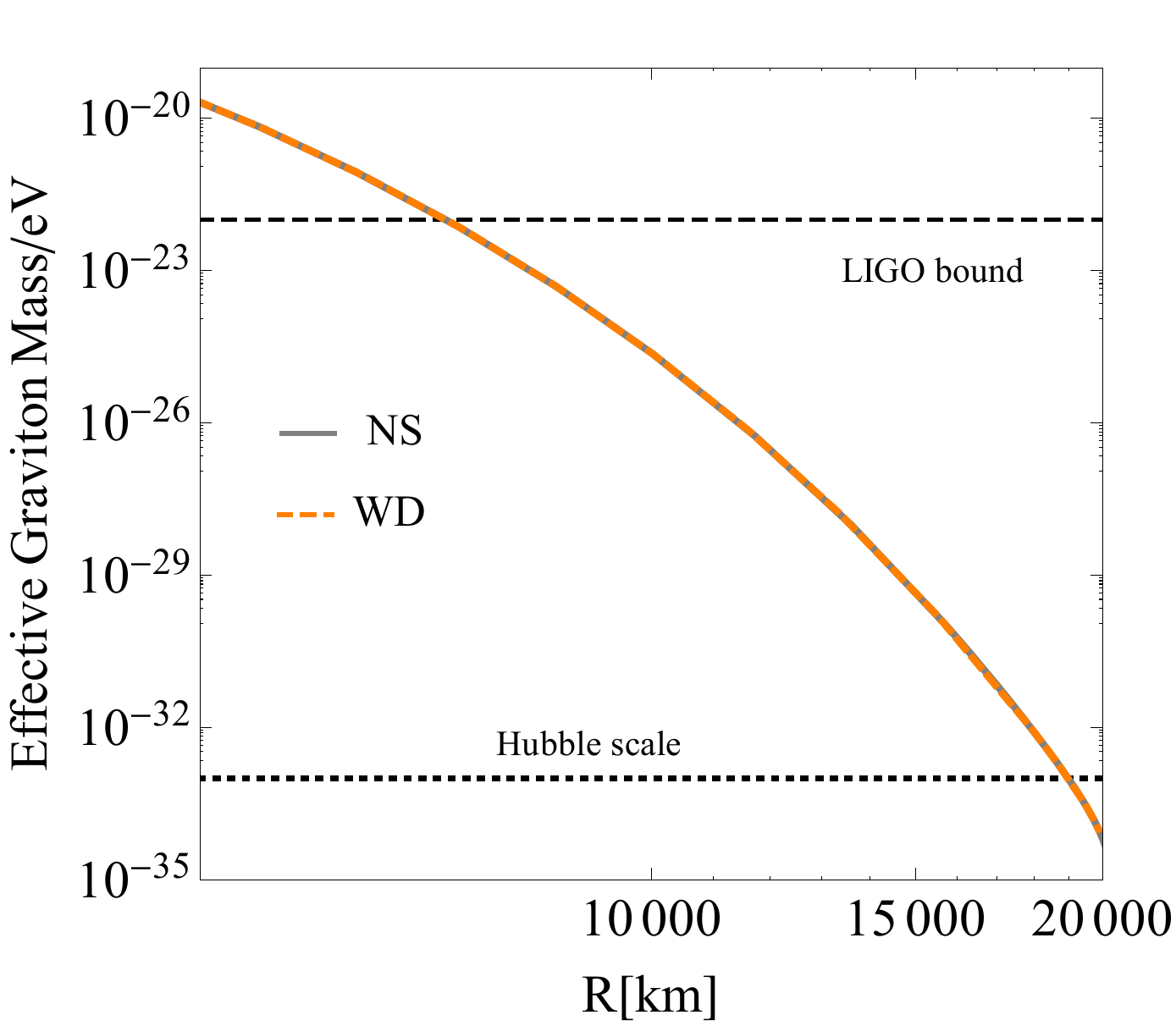}
\caption{Effective graviton mass with respect to the distance from the center of the star. We choose $m=0.1{\rm km}^{-1}$, $m_\si=0.001{\rm km}^{-1}$, and other parameters are as the same as Fig.~\ref{fig:nswd}. NS labels for a neutron star solution and  WD for a white dwarf solution. For a typical star solution, the effective graviton mass decreases to the Hubble scale at around 19000km, which is well within the solar radius, so the star solution trivially pass all the graviton mass bounds and the weak gravity tests such as the solar system gravity tests or the binary pulsar gravity tests. Yet, for the same set of parameters, the relativistic star solution can deviation from general relativity by about 10\% in the strong gravity regimes.}
\label{fig:gravimass}
\end{figure}

Finally, we would like to emphasize a point we briefly mentioned in the Introduction. One salient and interesting feature of the star as well as the black hole solutions in MVMG is that the effective graviton mass in the theory is very large near the center of the star or black hole but decreases very rapidly to the cosmological value away from the star or black hole. As we mentioned in Section \ref{sec:mvmg}, we let the effective graviton mass squared $V(\si)$ contain a small $m_0^2$, where $m_0$ is of the current Hubble scale, so when $\si$ goes to zero in weak gravity regimes, the theory effectively goes back to the original dRGT model. From Fig.~\ref{fig:gravimass}, we can see that the graviton mass $V^{1/2}$ decreases (faster than) exponentially to the Hubble scale away from the center for a typical neutron star and a typical white dwarf. This behavior is independent of the equation of state of the star, as we can see that for both the neutron star and white dwarf solution in Fig.~\ref{fig:gravimass}, despite quite different equations of state, the graviton mass reduces almost identically to the cosmological background value $m_0$ at around 19000km from the center, which is well within the radius of the sun.

The solar system gravity tests significantly constrain deviations from general relativity in the weak gravity regimes. In particular, the Yukawa force and fifth force tests in the solar system set the graviton mass bound to be close to the current Hubble scale if specific massive gravity models are assumed or some other assumptions are made, but the mass bounds are significantly weaker for model independent tests or if only focusing on the more reliable tests \cite{deRham:2016nuf}. In any case, the solar system gravity constraints are valid at distances greater than the solar radius. The binary pulsar gravity tests also constrain deviations from general relativity within about a percentage level, but they are also valid at distances about one solar radius. There are also other bounds on the graviton mass obtained by, for example, testing the dispersion relation of the graviton propagation across long distances. The aLIGO graviton mass bound is one of this kind, which sets the graviton mass to be 10 orders of greater than the current Hubble scale. The aLIGO bound is a very reliable bound, but it is much weaker than some other bounds \cite{deRham:2016nuf}. 

The original dRGT model satisfies all these bounds if a Hubble scale graviton mass is assumed. Therefore, the typical solutions in Fig.~\ref{fig:gravimass} trivially pass all these weak gravity tests and all the current bounds on the graviton mass. Yet, for these same solutions in the strong gravity regime, the deviations from general relativity are around 8\% to 10\%, as can be seen in Fig.~\ref{fig:nswd}. This exemplifies a central feature of MVMG, which is that the model can easily satisfy the weak gravity tests and at the same time can give rise to testable deviations from general relativity in the strong gravity regime.

\section{Summary}
\label{conclu}

In this paper, we have investigated the compact star solutions in mass-varying massive gravity. In the conventional weak gravity or cosmological environments, the mass of the graviton has been constrained to be very small, and if the graviton mass is of the current Hubble scale, the massive graviton effects may account for the observed late time cosmic acceleration \cite{deRham:2016nuf}. However, for such a small graviton mass, its effects on compact stars is negligibly small. In mass-varying massive gravity, the graviton mass can vary in different environments, as determined by a scalar field which takes different values in different gravity regimes. We have shown that the graviton mass can be extremely large near the compact stars while quickly decreasing to its cosmological value away from the stars. This is exactly the same as the case of black holes in MVMG, which has been previously investigated \cite{Zhang:2017jze}. The effects from the massive graviton and the environmental scalar field can be very small or significantly affect the mass-radius relation of the compact star, depending on the choice of the parameters in the theory.  Also, we have surveyed the MVMG parameter space ($m,m_\si$) for deviations from general relativity in terms of the mass of the compact star. We see that neutron stars can constrain MVMG better than white dwarfs for larger $m$ and $m_\si$, while white dwarfs have a slight advantage for smaller $m$ and larger $m_\si$.

We have found that there typically exists a discrete tower of relativistic star solutions for a given central matter energy density. In this tower, the star solution with the highest mass corresponds to an environmental scalar profile that falls directly to zero, while the scalar field oscillates to zero for the solutions with lower masses, exactly one solution for each possible oscillation pattern. We have also found that the number of possible solutions in the tower depends on the parameter $m$ and $m_\si$ in the theory.

~\\

    \noindent{\bf Acknowledgments}:
We would like to thank Qing-Guo Huang, Fan Wang and Jun Zhang for helpful suggestions and discussions. SYZ acknowledges support from the starting grant from University of Science and Technology of China under grant No.~KY2030000089 and is also supported by National Natural Science Foundation of China under grant No.~GG2030040375 and grant No.~11947301.\\

~\\

\bibliographystyle{JHEP}
\bibliography{refs}

\providecommand{\href}[2]{#2}\begingroup\raggedright\begin{thebibliography}{10}

\bibitem{Abbott:2016blz}
{\scshape LIGO Scientific, Virgo} collaboration, B.~P. Abbott et~al.,
  \emph{{Observation of Gravitational Waves from a Binary Black Hole Merger}},
  \href{http://dx.doi.org/10.1103/PhysRevLett.116.061102}{\emph{Phys. Rev.
  Lett.} {\bfseries 116} (2016) 061102},
  [\href{https://arxiv.org/abs/1602.03837}{{\ttfamily 1602.03837}}].

\bibitem{TheLIGOScientific:2017qsa}
{\scshape LIGO Scientific, Virgo} collaboration, B.~P. Abbott et~al.,
  \emph{{GW170817: Observation of Gravitational Waves from a Binary Neutron
  Star Inspiral}},
  \href{http://dx.doi.org/10.1103/PhysRevLett.119.161101}{\emph{Phys. Rev.
  Lett.} {\bfseries 119} (2017) 161101},
  [\href{https://arxiv.org/abs/1710.05832}{{\ttfamily 1710.05832}}].

\bibitem{deRham:2010ik}
C.~de~Rham and G.~Gabadadze, \emph{{Generalization of the Fierz-Pauli Action}},
  \href{http://dx.doi.org/10.1103/PhysRevD.82.044020}{\emph{Phys. Rev.}
  {\bfseries D82} (2010) 044020},
  [\href{https://arxiv.org/abs/1007.0443}{{\ttfamily 1007.0443}}].

\bibitem{deRham:2010kj}
C.~de~Rham, G.~Gabadadze and A.~J. Tolley, \emph{{Resummation of Massive
  Gravity}},
  \href{http://dx.doi.org/10.1103/PhysRevLett.106.231101}{\emph{Phys. Rev.
  Lett.} {\bfseries 106} (2011) 231101},
  [\href{https://arxiv.org/abs/1011.1232}{{\ttfamily 1011.1232}}].

\bibitem{Hassan:2011hr}
S.~F. Hassan and R.~A. Rosen, \emph{{Resolving the Ghost Problem in non-Linear
  Massive Gravity}},
  \href{http://dx.doi.org/10.1103/PhysRevLett.108.041101}{\emph{Phys. Rev.
  Lett.} {\bfseries 108} (2012) 041101},
  [\href{https://arxiv.org/abs/1106.3344}{{\ttfamily 1106.3344}}].

\bibitem{Hassan:2011ea}
S.~F. Hassan and R.~A. Rosen, \emph{{Confirmation of the Secondary Constraint
  and Absence of Ghost in Massive Gravity and Bimetric Gravity}},
  \href{http://dx.doi.org/10.1007/JHEP04(2012)123}{\emph{JHEP} {\bfseries 04}
  (2012) 123}, [\href{https://arxiv.org/abs/1111.2070}{{\ttfamily 1111.2070}}].

\bibitem{deRham:2014zqa}
C.~de~Rham, \emph{{Massive Gravity}},
  \href{http://dx.doi.org/10.12942/lrr-2014-7}{\emph{Living Rev. Rel.}
  {\bfseries 17} (2014) 7}, [\href{https://arxiv.org/abs/1401.4173}{{\ttfamily
  1401.4173}}].

\bibitem{Schmidt-May:2015vnx}
A.~Schmidt-May and M.~von Strauss, \emph{{Recent developments in bimetric
  theory}}, \href{http://dx.doi.org/10.1088/1751-8113/49/18/183001}{\emph{J.
  Phys.} {\bfseries A49} (2016) 183001},
  [\href{https://arxiv.org/abs/1512.00021}{{\ttfamily 1512.00021}}].

\bibitem{Hinterbichler:2011tt}
K.~Hinterbichler, \emph{{Theoretical Aspects of Massive Gravity}},
  \href{http://dx.doi.org/10.1103/RevModPhys.84.671}{\emph{Rev. Mod. Phys.}
  {\bfseries 84} (2012) 671--710},
  [\href{https://arxiv.org/abs/1105.3735}{{\ttfamily 1105.3735}}].

\bibitem{Boulware:1973my}
D.~G. Boulware and S.~Deser, \emph{{Can gravitation have a finite range?}},
  \href{http://dx.doi.org/10.1103/PhysRevD.6.3368}{\emph{Phys. Rev.} {\bfseries
  D6} (1972) 3368--3382}.

\bibitem{deRham:2016nuf}
C.~de~Rham, J.~T. Deskins, A.~J. Tolley and S.-Y. Zhou, \emph{{Graviton Mass
  Bounds}}, \href{http://dx.doi.org/10.1103/RevModPhys.89.025004}{\emph{Rev.
  Mod. Phys.} {\bfseries 89} (2017) 025004},
  [\href{https://arxiv.org/abs/1606.08462}{{\ttfamily 1606.08462}}].

\bibitem{Riess:1998cb}
{\scshape Supernova Search Team} collaboration, A.~G. Riess et~al.,
  \emph{{Observational evidence from supernovae for an accelerating universe
  and a cosmological constant}},
  \href{http://dx.doi.org/10.1086/300499}{\emph{Astron. J.} {\bfseries 116}
  (1998) 1009--1038}, [\href{https://arxiv.org/abs/astro-ph/9805201}{{\ttfamily
  astro-ph/9805201}}].

\bibitem{Perlmutter:1998np}
{\scshape Supernova Cosmology Project} collaboration, S.~Perlmutter et~al.,
  \emph{{Measurements of $\Omega$ and $\Lambda$ from 42 high redshift
  supernovae}}, \href{http://dx.doi.org/10.1086/307221}{\emph{Astrophys. J.}
  {\bfseries 517} (1999) 565--586},
  [\href{https://arxiv.org/abs/astro-ph/9812133}{{\ttfamily
  astro-ph/9812133}}].

\bibitem{Huang:2012pe}
Q.-G. Huang, Y.-S. Piao and S.-Y. Zhou, \emph{{Mass-Varying Massive Gravity}},
  \href{http://dx.doi.org/10.1103/PhysRevD.86.124014}{\emph{Phys. Rev.}
  {\bfseries D86} (2012) 124014},
  [\href{https://arxiv.org/abs/1206.5678}{{\ttfamily 1206.5678}}].

\bibitem{Huang:2013mha}
Q.-G. Huang, K.-C. Zhang and S.-Y. Zhou, \emph{{Generalized massive gravity in
  arbitrary dimensions and its Hamiltonian formulation}},
  \href{http://dx.doi.org/10.1088/1475-7516/2013/08/050}{\emph{JCAP} {\bfseries
  1308} (2013) 050}, [\href{https://arxiv.org/abs/1306.4740}{{\ttfamily
  1306.4740}}].

\bibitem{Zhang:2017jze}
J.~Zhang and S.-Y. Zhou, \emph{{Can the graviton have a large mass near black
  holes?}}, \href{http://dx.doi.org/10.1103/PhysRevD.97.081501}{\emph{Phys.
  Rev.} {\bfseries D97} (2018) 081501},
  [\href{https://arxiv.org/abs/1709.07503}{{\ttfamily 1709.07503}}].

\bibitem{Tolley:2015ywa}
A.~J. Tolley, D.-J. Wu and S.-Y. Zhou, \emph{{Hairy black holes in scalar
  extended massive gravity}},
  \href{http://dx.doi.org/10.1103/PhysRevD.92.124063}{\emph{Phys. Rev.}
  {\bfseries D92} (2015) 124063},
  [\href{https://arxiv.org/abs/1510.05208}{{\ttfamily 1510.05208}}].

\bibitem{Cardoso:2016rao}
V.~Cardoso, E.~Franzin and P.~Pani, \emph{{Is the gravitational-wave ringdown a
  probe of the event horizon?}},
  \href{http://dx.doi.org/10.1103/PhysRevLett.117.089902,
  10.1103/PhysRevLett.116.171101}{\emph{Phys. Rev. Lett.} {\bfseries 116}
  (2016) 171101}, [\href{https://arxiv.org/abs/1602.07309}{{\ttfamily
  1602.07309}}].

\bibitem{Cardoso:2017cqb}
V.~Cardoso and P.~Pani, \emph{{Tests for the existence of black holes through
  gravitational wave echoes}},
  \href{http://dx.doi.org/10.1038/s41550-017-0225-y}{\emph{Nat. Astron.}
  {\bfseries 1} (2017) 586--591},
  [\href{https://arxiv.org/abs/1709.01525}{{\ttfamily 1709.01525}}].

\bibitem{Katsuragawa:2015lbl}
T.~Katsuragawa, S.~Nojiri, S.~D. Odintsov and M.~Yamazaki, \emph{{Relativistic
  stars in de Rham-Gabadadze-Tolley massive gravity}},
  \href{http://dx.doi.org/10.1103/PhysRevD.93.124013}{\emph{Phys. Rev.}
  {\bfseries D93} (2016) 124013},
  [\href{https://arxiv.org/abs/1512.00660}{{\ttfamily 1512.00660}}].

\bibitem{Sullivan:2017kwo}
A.~Sullivan and N.~Yunes, \emph{{Slowly-Rotating Neutron Stars in Massive
  Bigravity}}, \href{http://dx.doi.org/10.1088/1361-6382/aaa3ab}{\emph{Class.
  Quant. Grav.} {\bfseries 35} (2018) 045003},
  [\href{https://arxiv.org/abs/1709.03311}{{\ttfamily 1709.03311}}].

\bibitem{Hendi:2017ibm}
S.~H. Hendi, G.~H. Bordbar, B.~Eslam~Panah and S.~Panahiyan, \emph{{Neutron
  stars structure in the context of massive gravity}},
  \href{http://dx.doi.org/10.1088/1475-7516/2017/07/004}{\emph{JCAP} {\bfseries
  1707} (2017) 004}, [\href{https://arxiv.org/abs/1701.01039}{{\ttfamily
  1701.01039}}].

\bibitem{EslamPanah:2018evk}
B.~Eslam~Panah and H.~L. Liu, \emph{{White dwarfs in de Rham-Gabadadze-Tolley
  like massive gravity}},
  \href{http://dx.doi.org/10.1103/PhysRevD.99.104074}{\emph{Phys. Rev.}
  {\bfseries D99} (2019) 104074},
  [\href{https://arxiv.org/abs/1805.10650}{{\ttfamily 1805.10650}}].

\bibitem{Akmal:1998cf}
A.~Akmal, V.~R. Pandharipande and D.~G. Ravenhall, \emph{{The Equation of state
  of nucleon matter and neutron star structure}},
  \href{http://dx.doi.org/10.1103/PhysRevC.58.1804}{\emph{Phys. Rev.}
  {\bfseries C58} (1998) 1804--1828},
  [\href{https://arxiv.org/abs/nucl-th/9804027}{{\ttfamily nucl-th/9804027}}].

\bibitem{Douchin:2001sv}
F.~Douchin and P.~Haensel, \emph{{A unified equation of state of dense matter
  and neutron star structure}},
  \href{http://dx.doi.org/10.1051/0004-6361:20011402}{\emph{Astron. Astrophys.}
  {\bfseries 380} (2001) 151},
  [\href{https://arxiv.org/abs/astro-ph/0111092}{{\ttfamily
  astro-ph/0111092}}].

\bibitem{Shen:1998by}
H.~Shen, H.~Toki, K.~Oyamatsu and K.~Sumiyoshi, \emph{{Relativistic equation of
  state of nuclear matter for supernova explosion}},
  \href{http://dx.doi.org/10.1143/PTP.100.1013}{\emph{Prog. Theor. Phys.}
  {\bfseries 100} (1998) 1013},
  [\href{https://arxiv.org/abs/nucl-th/9806095}{{\ttfamily nucl-th/9806095}}].

\bibitem{Demorest:2010bx}
P.~Demorest, T.~Pennucci, S.~Ransom, M.~Roberts and J.~Hessels, \emph{{Shapiro
  Delay Measurement of A Two Solar Mass Neutron Star}},
  \href{http://dx.doi.org/10.1038/nature09466}{\emph{Nature} {\bfseries 467}
  (2010) 1081--1083}, [\href{https://arxiv.org/abs/1010.5788}{{\ttfamily
  1010.5788}}].

\bibitem{Antoniadis:2013pzd}
J.~Antoniadis et~al., \emph{{A Massive Pulsar in a Compact Relativistic
  Binary}}, \href{http://dx.doi.org/10.1126/science.1233232}{\emph{Science}
  {\bfseries 340} (2013) 6131},
  [\href{https://arxiv.org/abs/1304.6875}{{\ttfamily 1304.6875}}].

\bibitem{Haensel:2004nu}
P.~Haensel and A.~Y. Potekhin, \emph{{Analytical representations of unified
  equations of state of neutron-star matter}},
  \href{http://dx.doi.org/10.1051/0004-6361:20041722}{\emph{Astron. Astrophys.}
  {\bfseries 428} (2004) 191--197},
  [\href{https://arxiv.org/abs/astro-ph/0408324}{{\ttfamily
  astro-ph/0408324}}].

\bibitem{Will:2014kxa}
C.~M. Will, \emph{{The Confrontation between General Relativity and
  Experiment}}, \href{http://dx.doi.org/10.12942/lrr-2014-4}{\emph{Living Rev.
  Rel.} {\bfseries 17} (2014) 4},
  [\href{https://arxiv.org/abs/1403.7377}{{\ttfamily 1403.7377}}].

\bibitem{Chandrasekhar:1931ih}
S.~Chandrasekhar, \emph{{The maximum mass of ideal white dwarfs}},
  \href{http://dx.doi.org/10.1086/143324}{\emph{Astrophys. J.} {\bfseries 74}
  (1931) 81--82}.

\bibitem{Chandrasekhar:1935zz}
S.~Chandrasekhar, \emph{{The highly collapsed configurations of a stellar mass
  (Second paper)}}, \href{http://dx.doi.org/10.1093/mnras/95.3.207}{\emph{Mon.
  Not. Roy. Astron. Soc.} {\bfseries 95} (1935) 207--225}.

\bibitem{Howell:2006vn}
{\scshape SNLS} collaboration, D.~A. Howell et~al., \emph{{The type Ia
  supernova SNLS-03D3bb from a super-Chandrasekhar-mass white dwarf star}},
  \href{http://dx.doi.org/10.1038/nature05103}{\emph{Nature} {\bfseries 443}
  (2006) 308}, [\href{https://arxiv.org/abs/astro-ph/0609616}{{\ttfamily
  astro-ph/0609616}}].

\bibitem{Scalzo:2010xd}
R.~A. Scalzo et~al., \emph{{Nearby Supernova Factory Observations of SN 2007if:
  First Total Mass Measurement of a Super-Chandrasekhar-Mass Progenitor}},
  \href{http://dx.doi.org/10.1088/0004-637X/713/2/1073}{\emph{Astrophys. J.}
  {\bfseries 713} (2010) 1073--1094},
  [\href{https://arxiv.org/abs/1003.2217}{{\ttfamily 1003.2217}}].

\bibitem{Hicken:2007ap}
M.~Hicken, P.~M. Garnavich, J.~L. Prieto, S.~Blondin, D.~L. DePoy, R.~P.
  Kirshner et~al., \emph{{The Luminous and Carbon-Rich Supernova 2006gz: A
  Double Degenerate Merger?}},
  \href{http://dx.doi.org/10.1086/523301}{\emph{Astrophys. J.} {\bfseries 669}
  (2007) L17--L20}, [\href{https://arxiv.org/abs/0709.1501}{{\ttfamily
  0709.1501}}].

\bibitem{Yamanaka:2009dp}
M.~Yamanaka et~al., \emph{{Early phase observations of extremely luminous Type
  Ia Supernova 2009dc}},
  \href{http://dx.doi.org/10.1088/0004-637X/707/2/L118}{\emph{Astrophys. J.}
  {\bfseries 707} (2009) L118--L122},
  [\href{https://arxiv.org/abs/0908.2059}{{\ttfamily 0908.2059}}].

\bibitem{Silverman:2010bh}
J.~M. Silverman, M.~Ganeshalingam, W.~Li, A.~V. Filippenko, A.~A. Miller and
  D.~Poznanski, \emph{{Fourteen Months of Observations of the Possible
  Super-Chandrasekhar Mass Type Ia Supernova 2009dc}},
  \href{http://dx.doi.org/10.1111/j.1365-2966.2010.17474.x}{\emph{Mon. Not.
  Roy. Astron. Soc.} {\bfseries 410} (2011) 585},
  [\href{https://arxiv.org/abs/1003.2417}{{\ttfamily 1003.2417}}].

\bibitem{Taubenberger:2010qv}
S.~Taubenberger et~al., \emph{{High luminosity, slow ejecta and persistent
  carbon lines: SN 2009dc challenges thermonuclear explosion scenarios}},
  \href{http://dx.doi.org/10.1111/j.1365-2966.2010.18107.x}{\emph{Mon. Not.
  Roy. Astron. Soc.} {\bfseries 412} (2011) 2735},
  [\href{https://arxiv.org/abs/1011.5665}{{\ttfamily 1011.5665}}].

\end{thebibliography}\endgroup

\end{document}